 \documentclass{jaa}
\usepackage[flushleft]{threeparttable}
\usepackage{textcomp}
\usepackage{gensymb}
\usepackage{natbib}
\usepackage{graphicx}
\usepackage[dvipsnames]{xcolor}
\usepackage{subcaption}
\definecolor{xlinkcolor}{cmyk}{1,1,0,0}
 \usepackage[bookmarks=false,         
     pdfnewwindow=true,      
     colorlinks=true,    
     linkcolor=xlinkcolor,     
     citecolor=xlinkcolor,     
     filecolor=xlinkcolor,  
     urlcolor=xlinkcolor,      
final=true
 ]{hyperref}

\begin{document} \sloppy

\title{Imaging Calibration of AstroSat Cadmium Zinc Telluride Imager (CZTI)}

\title{Imaging Calibration of AstroSat Cadmium Zinc Telluride Imager (CZTI)}

\author{Ajay Vibhute\textsuperscript{1,2,*}, Dipankar Bhattacharya\textsuperscript{2}, N. P. S. Mithun\textsuperscript{3}, V. Bhalerao\textsuperscript{4}, A. R. Rao\textsuperscript{5}, S.V. Vadawale\textsuperscript{3} }


\affilOne{\textsuperscript{1}Savitribai Phule Pune University, Pune, Maharashtra, India\\}
\affilTwo{\textsuperscript{2}Inter-University Centre for Astronomy and Astrophysics (IUCAA), Post Bag 4, Ganeshkhind, Pune, India\\}
\affilThree{\textsuperscript{3}Physical Research Laboratory, Ahmedabad, Gujarat, India\\}
\affilFour{\textsuperscript{4}Indian Institute of Technology, Bombay, India\\}
\affilFive{\textsuperscript{5}Tata Institute of Fundamental Research, Homi Bhabha Road, Mumbai, India\\}


\twocolumn[{
\maketitle
\corres{ajay@iucaa.in}
\msinfo{15 October 2020}{15 October 2020}

\begin{abstract}
AstroSat is India's first space-based astronomical observatory, launched on September 28, 2015. One of the payloads aboard AstroSat is the Cadmium Zinc Telluride Imager (CZTI), operating at hard X-rays. CZTI employs a two-dimensional coded aperture mask for the purpose of imaging. In this paper, we discuss various image reconstruction algorithms adopted for the test and calibration of the imaging capability of CZTI and present results from CZTI on-ground as well as in-orbit image calibration.
\end{abstract}
}
]

\section{Introduction}
AstroSat~\citep{astrosat}, India's first dedicated space astronomy observatory, covers a broad energy band including optical, Ultra-Violet (UV) and soft to hard X-rays, using four different co-pointed payloads, namely Ultra-Violet Imaging Telescope [UVIT; \citealt{uvit_tandon_2017}], Soft X-ray Telescope [SXT; \citealt{singh2017sxt}],  Large Area X-ray Proportional Counters [LAXPC; \citealt{yadavlaxpc2016}], and Cadmium Zinc Telluride Imager [CZTI; \citealt{vadale2016inorbitperformance}]. In addition; a Scanning Sky Monitor [SSM; \citealt{ramadevi2017ssm}]  is mounted perpendicular to other instruments and operates independently to search for X-ray transients.  The SXT carries out x-ray imaging of the sky in the soft X-ray band (0.3~keV - 8~keV) using reflecting optics and CZTI in the hard X-ray band (20~keV-100keV) using a Coded Mask.

In this paper section~\ref{sec_intoczti} gives an introduction to the AstroSat CZT Imager, and section~\ref{czticalibration} presents results from the on-ground and in-flight calibration.

\subsection{Cadmium Zinc Telluride Imager}\label{sec_intoczti}
	The CZT detector is sensitive up to 500$~keV$, and CZTI electronics are tuned to discriminate photons up to 250 $~keV$. Coded Mask imaging~\citep{skinner1984} with the CZTI is designed to operate in the energy range  20 to 100$~keV$. 
\begin{figure}[ht]
    \centering
     \includegraphics[width=0.45\textwidth]{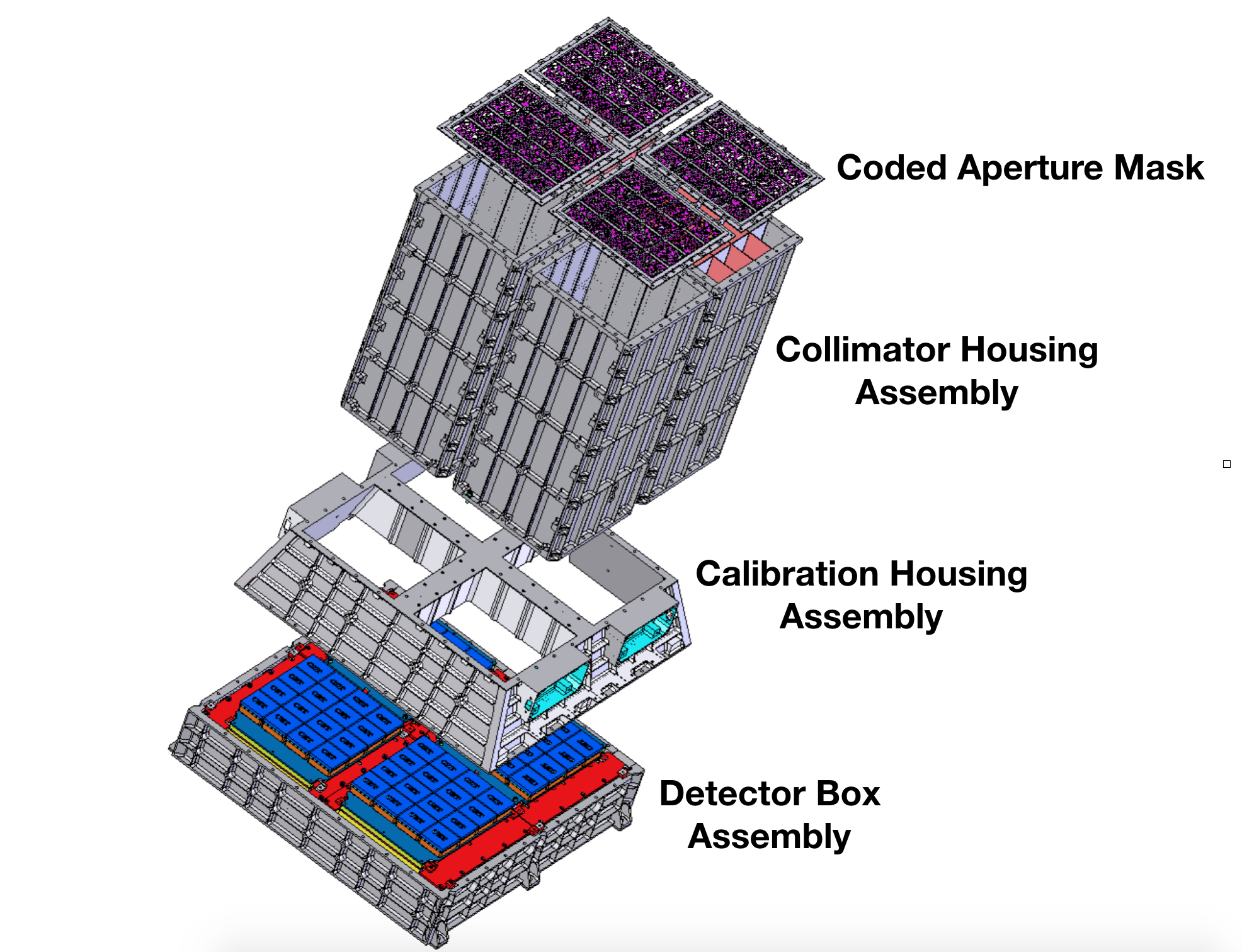}
        \caption{Assembly of the CZTI}
       \label{fig_czti_assembly} 
\end{figure}

\begin{figure*}[ht!]
    \centering
    \includegraphics[width=0.7\textwidth]{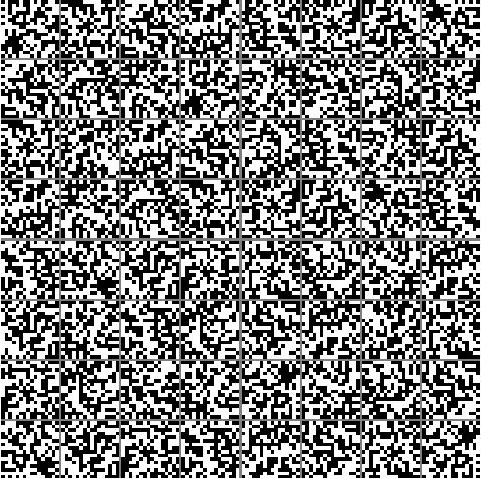}
    \caption{CZTI coded aperture mask for all four quadrants designed using 255 element pseudo-noise Hadamard set Uniformly Redundant Arrays. One extra closed element is added to each URA to obtain a square pattern. Here black areas represent closed mask elements and white areas represent open ones.}
    \label{fig_czti_fullmask} 
\end{figure*}

	CZTI is a 'box-type' or 'simple' coded mask telescope where the size of the coded mask and the detector are the same. The full CZTI detector is illuminated only by the on-axis source. Off-axis sources illuminate only a part of the detector depending on the position of the source, this is called ``partial coding''. The coded mask for CZTI is designed with seven patterns based on 255-element pseudo-noise Hadamard set Uniformly Redundant Arrays (URA)~\citep{caroli}. Each pattern has 16$\times$16 mask elements and used as a mask for an individual detector module. A random arrangement of these patterns into a 4 $\times$ 4 array results in the mask pattern for the first quadrant (quadrant A). The coded masks for the other quadrants (B, C, and D) were obtained by rotating the mask pattern of quadrant A by 90$\degree$, 180$\degree$ and 270$\degree$ respectively. Figure~\ref{fig_czti_fullmask} shows the mask pattern for all four quadrants of CZTI. In the figure, closed mask elements are represented in black and the open ones in white.

CZTI has a total geometric area of 976~cm$^2$ achieved using 64 detector modules, divided across four independent quadrants. Each CZT detector module is of size 39.06~mm$\times$39.06~mm$\times$ 5~mm and contains 256 pixels arranged in a 16$\times$16 array.  An individual pixel in the array is of size 2.46~mm$\times$2.46~mm, except at edge rows which are 2.31~mm wide. Thus, the edge pixels are of size 2.31~mm$\times$2.46~mm, and the corner pixels are of size 2.31~mm$\times$2.31~mm. 

A passive collimator wall of height 400~mm separates any two adjacent modules, restricting the view of each detector module to the coded mask directly above it.  The collimator thus restricts the Field of View (FOV) of CZTI to 4.6$\degree$ $\times$ 4.6$\degree$ FWHM at energies below 100~keV. For energies above 100~keV, the collimator walls and the coded mask become progressively transparent, and allows the detection of Gamma-Ray Bursts from all over the sky \citep{rao2016grb}. Table~\ref{tab:cztchar} lists the main characteristics of CZTI. A detailed description of CZTI and its science goals may be found in \citep{bhalerao2017cadmium}.

\begin{table}
	\begin{tabular}{|p{3cm}|p{4.5cm}|}
    \hline
        Detector & Cadmium-Zinc-Telluride \\
    \hline
        Pixels &16384 (64 modules of 256 pixels each)\\
    \hline
        Pixel size & 2.46~mm $\times$ 2.46~mm \\
    \hline
        Imaging method & Coded Aperture Mask\\
    \hline
        Field of View & 4.6{\degree} $\times$  4.6{\degree}\\
    \hline
        Angular resolution& $\sim$ 8 arc min   (18 arc min geometric) \\
    \hline
        Energy resolution& $\sim$ 8\% @ 100 keV \\
    \hline
	Energy range &20  $-$ 100~keV for collimated FOV (20 $-$ 380~keV for all sky) \\
    \hline
        Sensitivity&0.5 mCrab (5 sigma; 10$^4$~$s$)\\
    \hline
\end{tabular}
\caption {The important characteristics of CZT Imager.}
 \label{tab:cztchar} 
\end{table}

		CZTI records the distribution of counts as a function of position on the detector. The observed count distribution does not correspond to a sky image. The latter is obtained by subjecting the observed pattern to a image reconstruction procedure.

\section{Imaging Calibration of CZTI}\label{czticalibration}
Like every space astronomy payload, the CZTI flight model underwent a phase of calibration on ground to characterise the
pixel behaviour, effective area, imaging and spectral response, etc. The ground-based calibration was carried out by exposing the instrument to various radioactive sources placed at a finite distance from the detector. After launch, the first six months were devoted to carry out in-flight calibration using observations of astronomical sources. 

CZTI image reconstruction is a two-step process, the first step is to acquire a spatially coded detector data, and the second step involves reconstruction of the observed image by decoding the collected data. The reconstruction is computationally expensive and performed offline, i.e., on the ground. A variety of image reconstruction algorithms are available, from which a choice is made based on the available computing resources, degree of crowding, nature of the background and the required accuracy. 
CZTI is a photon-counting detector and records the time of arrival, energy and position on the detector of each detected photon. CZTI image reconstruction starts by creating a two-dimensional map of total counts recorded in each pixel, called a Detector Plane Histogram (DPH). As the detection efficiency varies across the pixels, the DPH is normalized by the relative quantum efficiency (QE) of individual pixels to obtain a Detector Plane Image (DPI). The relative quantum efficiency (QE) of the pixels are available as a part of the CZTI Calibration Database (CALDB)\footnote{http://astrosat-ssc.iucaa.in/?q=cztiData}. The DPI represents a scaled shadow of the mask on the detector plane cast by the sources in the FOV. The pattern of the DPI depends on the position of the source in FOV and the recorded counts in each pixel depend on the intensities of these sources. If the position or intensity of a source changes, so would the total number of photons recorded by each pixel.  The DPI is used as an input to the image reconstruction algorithm. 
We have used  three variants of the Cross-Correlation, namely, mask cross-correlation~\citep{skinner_1987,codedmaskimaging}, shadow cross-correlation, balanced cross-correlation~\citep{fenimore_1978, codedmaskimaging} and RichardsonLucy~\citep{richardson,db_2006_rl,ravi_imaging_2003} techniques for CZTI imaging calibration.

In this paper we report the results of imaging calibration of CZTI, both on-ground and in-flight, in the sections to follow.
\subsection{CZTI On Ground Calibration}
We performed CZTI ground calibration in two phases, in the first phase, the coded mask and the imaging procedure were validated, and in the second phase, we calibrated the fully assembled CZTI to validate the imaging response.  

\subsubsection{On Ground Calibration using the Qualification Model}\label{ground_qm_cal}
\begin{figure}
    \centering
    \includegraphics[width=0.2\textwidth,height=0.3\textheight]{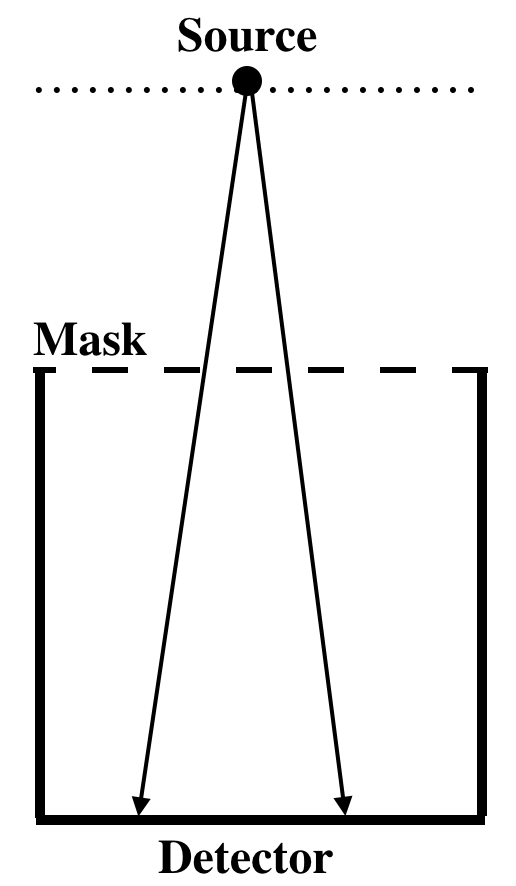}
    \caption[A schematic representation of the assembly made for CZTI ground calibration using Qualification Module]{ A schematic representation of the assembly made to test the coded mask imaging procedure using CZTI Qualification Module.  In this setup the source was placed 897.5~mm above the detector while the mask is located at a height of 484~mm above the detector.}
    \label{fig_test_jig}
\end{figure}

One detector module with 16$\times$16 pixels of a Qualification Model (QM) was used first to verify the imaging procedure. A fixture, schematically shown in figure~\ref{fig_test_jig},  was made to perform the ground calibration.  In the fixture, the source and the mask plate were placed at the height of $897.5~mm$ and $484~mm$ respectively above the detector. Two radioactive sources, Americium-241 (with a line at 59.54~keV) and Cobalt-57 (with lines at 122~keV and 136~keV) were used to validate the coded mask and the imaging procedure. 
	$^{241}$Am had count rate of 400 counts/sec and $^{57}$Co had 350 counts/sec. We selected several locations on the fixture, including the centre as well as a few offset locations and carried out multiple observations at these locations. The finite distance of the source from the detector causes the shadow to differ from what would be expected from a distant astronomical source.  In the laboratory setup built for CZTI ground calibration, one mask element cast its shadow over 2$\times$2 detector pixels. Hence, only a quarter of the mask illuminated the entire detector module, despite the mask and the detector being of the same physical size. For image reconstruction, we computed a library of the expected shadows of the mask on the detector, spanning source positions over $\pm 19.5$~mm from the centre in both X and Y directions, in steps of $0.25$~mm.

\begin{figure}
    \centering
    \includegraphics[width=0.5\textwidth]{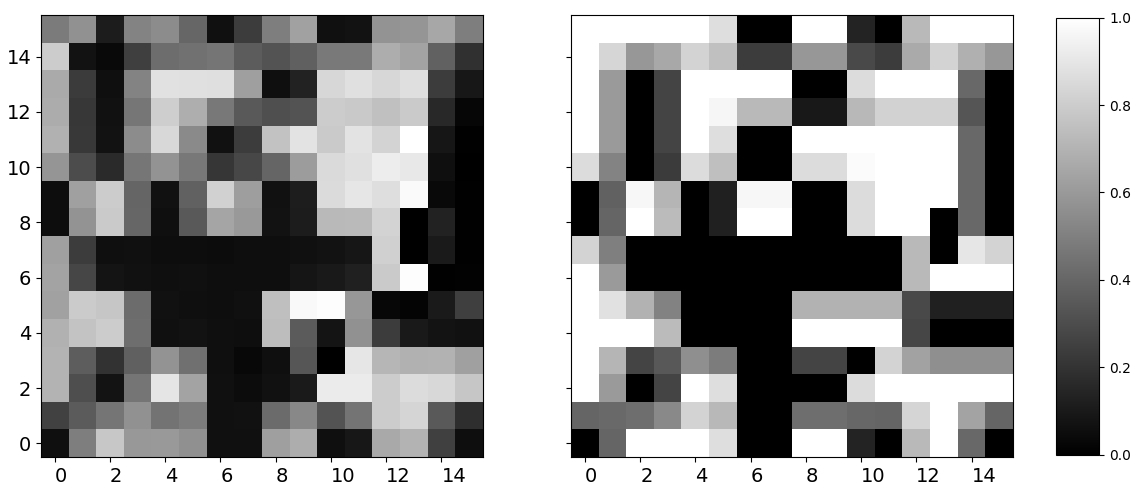}
	\caption{The expected (left panel) and observed (right panel) shadow of the mask on the detector when the radioactive source $^{241}$Am  was placed directly above the centre of the detector, in the arrangement shown in Fig.~\protect{\ref{fig_test_jig}}. The observed pattern is normalized and dead/noisy pixels are ignored. }

    \label{fig_ground_shadow}
\end{figure}

We obtained five different data sets by placing the $^{241}$Am source at different locations, including the centre and a few offset positions. One data set was also obtained by simultaneously placing two sources, $^{241}$Am and  $^{57}$Co at two different locations on the fixture. In figure~\ref{fig_ground_shadow}, the left panel shows the expected shadow pattern, and the right panel shows the observed pattern when the source was placed on a normal passing through the centre of the detector. The CZTI electronics recorded the pixel ID, energy and time of each detected photon, creating an event list.  The DPH were constructed by binning the events as a function of pixel number and along with the computed shadow library were used as input to the imaging procedure. For image reconstruction during the ground calibration, we used  Richardson-Lucy and cross-correlation algorithms. 

\begin{figure*}[ht!]
	\begin{subfigure}{0.45\textwidth}
    		\centering
     		\includegraphics[width=\textwidth]{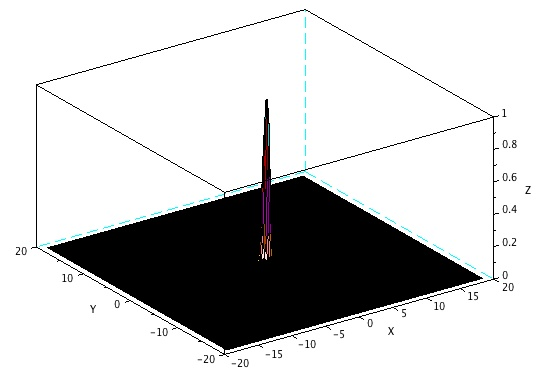}
		\caption{Reconstructed image using Richardson-Lucy}
	\end{subfigure}
	\begin{subfigure}{0.45\textwidth}
	    	\centering
		\includegraphics[width=\textwidth]{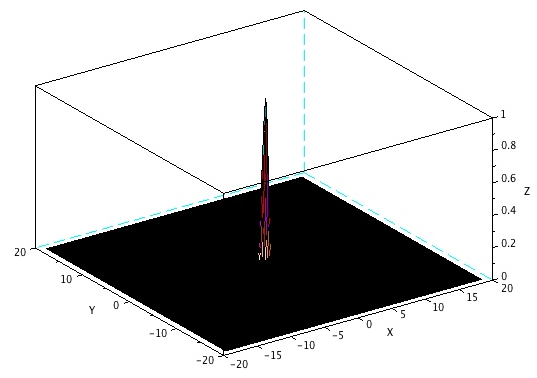} 
		\caption{Reconstructed image using  cross-correlation}
	\end{subfigure}
	 \caption{Reconstructed image using Richardson-Lucy and cross-correlation methods, when source is at the centre of the detector.}
	\label{fig_qm_image}	 

\end{figure*}

\begin{figure*}[ht!]
	\centering
	\begin{subfigure}{0.45\textwidth}
		\includegraphics[width=\textwidth]{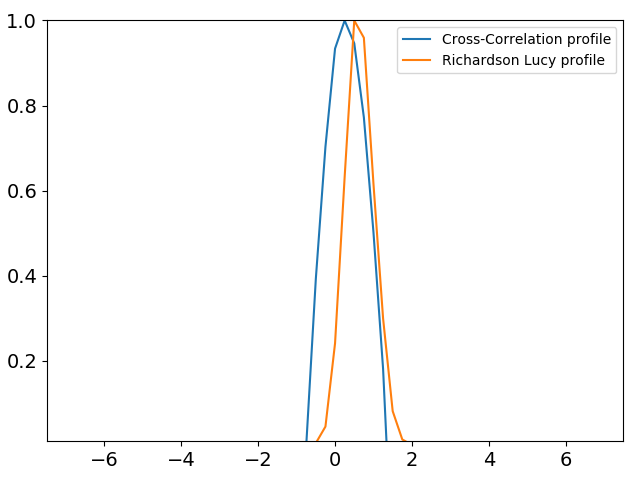}
		\caption{Image profile in X cross sections.}
	\end{subfigure}
	\begin{subfigure}{0.45\textwidth}
		\includegraphics[width=\textwidth]{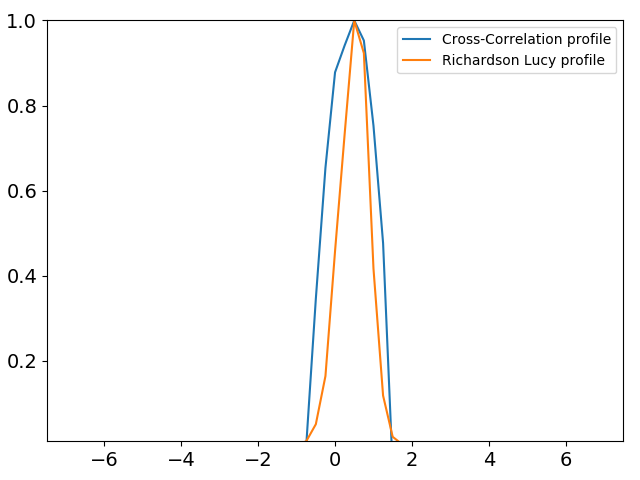}
		\caption{Image profile in Y cross section.}
	\end{subfigure}
	\caption{Image profiles of X and Y cross sections of the image reconstructed using Cross-Correlation and Richardson-Lucy algorithm.}
	\label{fig_rl_image_profiles}
\end{figure*}

\begin{figure}[ht!]
	\centering
     		\includegraphics[width=0.45\textwidth]{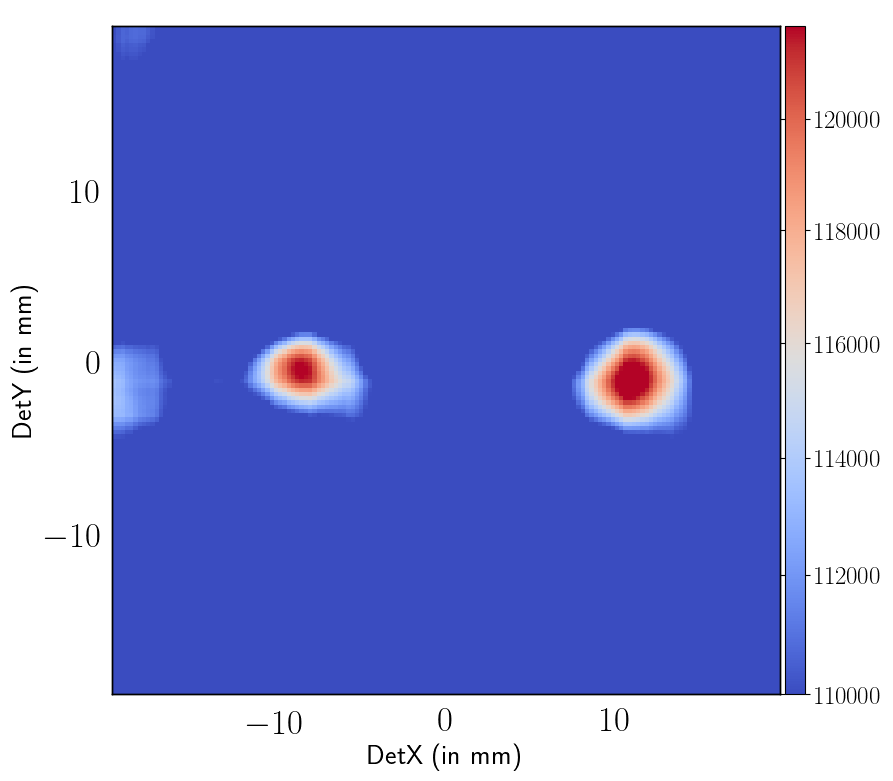}
		\caption{Reconstructed image using cross-correlation method for the simultaneous observation of two sources where $^{241}$Am source was placed at the offset of -10.25$~mm$ and $^{57}$Co was placed at the offset of +10.25$~mm$ from the center in X-direction. Sources were of almost equal strength.}
		\label{fig_qm_two_source_image}
\end{figure}

	Figure~\ref{fig_qm_image} shows the reconstructed image when $^{241}$Am source was placed at the center using Richardson-Lucy and cross-correlation methods, and image profiles of X and Y cross sections of the reconstructed image is shown in figure~\ref{fig_rl_image_profiles}.  Figure~\ref{fig_qm_two_source_image} shows the reconstructed image using cross-correlation for the simultaneous observation of two sources where $^{241}$Am source was placed at the offset of -10.25$~mm$ and $^{57}$Co was placed at the offset of -10.25$~mm$ from the center in X-direction.

We then compared the reconstructed source location from the image with the source position set in the lab during the acquisition of the data. Table~\ref{tab_qm_results} presents the results of this exercise.
\begin{table}
  \begin{center}

    \begin{tabular}{|c|c|c|}
        \hline
        Lab location (in mm) & CC (in mm)  &  RL (in mm)\\
      \hline
        (0.0,0.0) & (0.7,0.75) & (0.5,0.4)\\
        \hline
        (0.0,-10.25) & (0.2,-9.75) & (0.25,-9.7)\\
        \hline
        (0.0,10.25) & (0.5,11.0) & (0.5,11.0)\\
        \hline
        (-10.25,0.0) & (-9.85,0.5) & (-9.75,0.25)\\
        \hline
        (10.25,0.0) & (11.0,0.5) & (10.65,0.35)\\
        \hline
    \end{tabular}
  \end{center}
    \caption[Image reconstruction results from laboratory setup with one source] {Position of source in lab and reconstructed positions using  cross-correlation and Richardson-Lucy algorithm.}
    \label{tab_qm_results}
\end{table}
In the case of two-source observation, we constructed two separate DPHs, one from photons recorded near $60$~keV, and another from photons near $122$~keV. Table~\ref{tab_qm_results_twosrc} summarises the reconstruction result for the observation with two sources.

\begin{table*}[ht]
  \begin{center}

    \begin{tabular}{|c|c|c|c|}
    \hline
       Source & Lab location (in mm) & CC (in mm)  &  RL (in mm)\\
        
      \hline
      
        $^{241} Am$  &(-10.25,0.0) & (-9.0,0.0) & (-9.0,-0.2)\\
        \hline
        $^{57}Co$ &    (10.25,0.0) & (10.7,-1.0) & (10.5,-1.25)\\
    \hline
    \end{tabular}
  \end{center}
    \caption[Image reconstruction results from laboratory setup with two sources]{Lab position of the source during the observation and reconstructed source position using Cross-correlation and Richardson-Lucy algorithm.}
    \label{tab_qm_results_twosrc}
\end{table*}
The results from cross-correlation and Richardson-Lucy methods agreed within 0.5~mm. The source positions set in the lab and the reconstructed positions were consistent well within 1~mm, except for one case where the deviation was 1.25~mm. This discrepancy in the reconstructed position could be due to the inaccuracy in the manual placement of the radioactive source. The location accuracy of 1.0~mm at a source height of 897.5~mm corresponds to an angular accuracy of 3.8~arcmin. In all the above reconstructions, the total number of source photons used are more than a million and average background count was 350 counts/sec. To examine the reliability of the algorithm, image reconstruction was also performed with a smaller subset of photons. Event lists containing  $10^5$, $10^4$, $10^3$ and $10^2$ photons were selected randomly from the full event list and imaging was performed using them.  We repeated the imaging procedure with ten different random event sets in each case to estimate the variation in reconstruction results due to counting statistics.  Table~\ref{tab_lower_photons} lists the absolute difference (in mm) between the source location reconstructed using the subset event list and the full event list.
\begin{table*}[ht]
    \centering
    \begin{tabular}{|c|c|c|c|c|}
        \hline
        Trial No& $10^5 photons$ &  $10^4 photons$ &  $10^3 photons$ &  $10^2 photons$\\
        \hline
        1 & 0.0 & 0.0 & 0.0 & 0.55\\
        \hline
        2 & 0.0 & 0.25 & 2.70 & 0.55\\
        \hline
        3 & 0.0 & 0.25 & 0.35 & 0.55 \\
        \hline
        4 & 0.0 & 0.0 & 0.0 & 1.03\\
        \hline
        5 & 0.0 & 2.85 & 0.0 & 0.25\\
        \hline
        6 & 0.0 & 2.85 & 0.25 & 0.75\\
        \hline
        7 & 0.0 & 0.0  & 0.25  & 0.50\\
        \hline
        8 & 0.0 & 0.25 & 0.25 & 0.79\\
        \hline
        9 & 0.0 & 0.0 & 0.25 & 0.25\\
        \hline
        10 & 0.0 & 0.0 & 0.35 & 0.25\\
        \hline
    \end{tabular}
    \caption[Source location reconstruction with lower photons]{Absolute difference (in mm) between the source location using a smaller number of events and the full event list.}
    \label{tab_lower_photons}
\end{table*}
The results show that the reconstruction is quite robust, and the RMS deviation is within 1.3~mm (5~arcmin).

\subsubsection{On Ground Calibration using the Flight Module}\label{ground_fm_cal}

The results from the qualification model confirmed the working of the coded mask and the imaging procedure. The four quadrants of the CZTI Flight Model (FM) were then assembled and calibrated.

\begin{figure}
    \centering
     \includegraphics[width=0.5\textwidth]{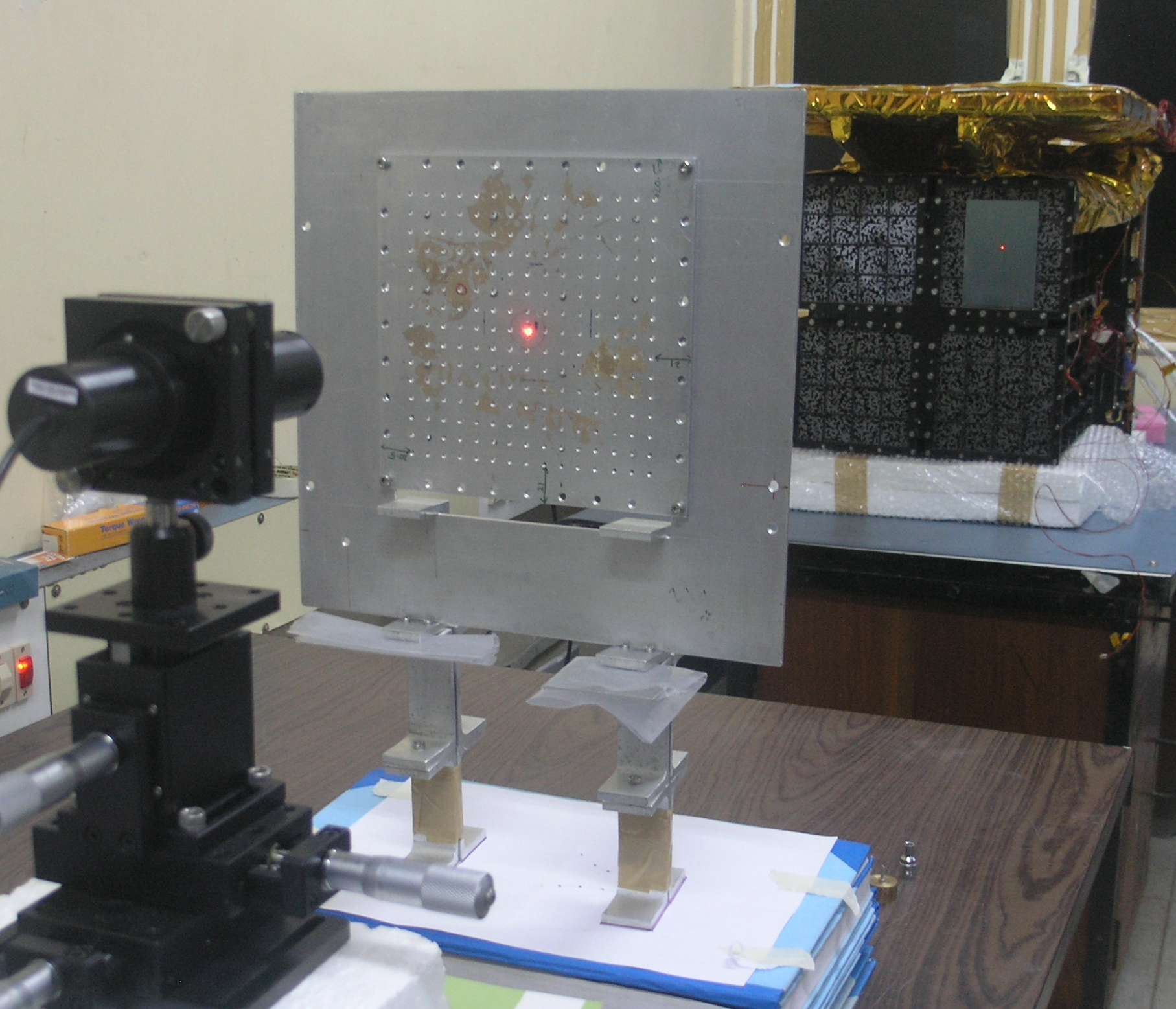}
    \caption[The laboratory setup used for FM calibration]{ Laboratory setup  at  Tata Institute of Fundamental Research (TIFR) used for the CZTI FM calibration}
       \label{fig_fm_lab_setup} 
\end{figure}

The ground calibration of the FM was carried out by shining three different radioactive sources on individual quadrants. In addition to $^{241}$Am and $^{57}$Co, a $^{109}$Cd source with lines at 22~keV and 88~keV was used. Figure~\ref{fig_fm_lab_setup} shows the laboratory setup used for the FM calibration. The perpendicularity of placement of the source with respect to the detector was ensured using a laser beam and a mirror.

\begin{figure}
    \centering
     \includegraphics[width=0.5\textwidth]{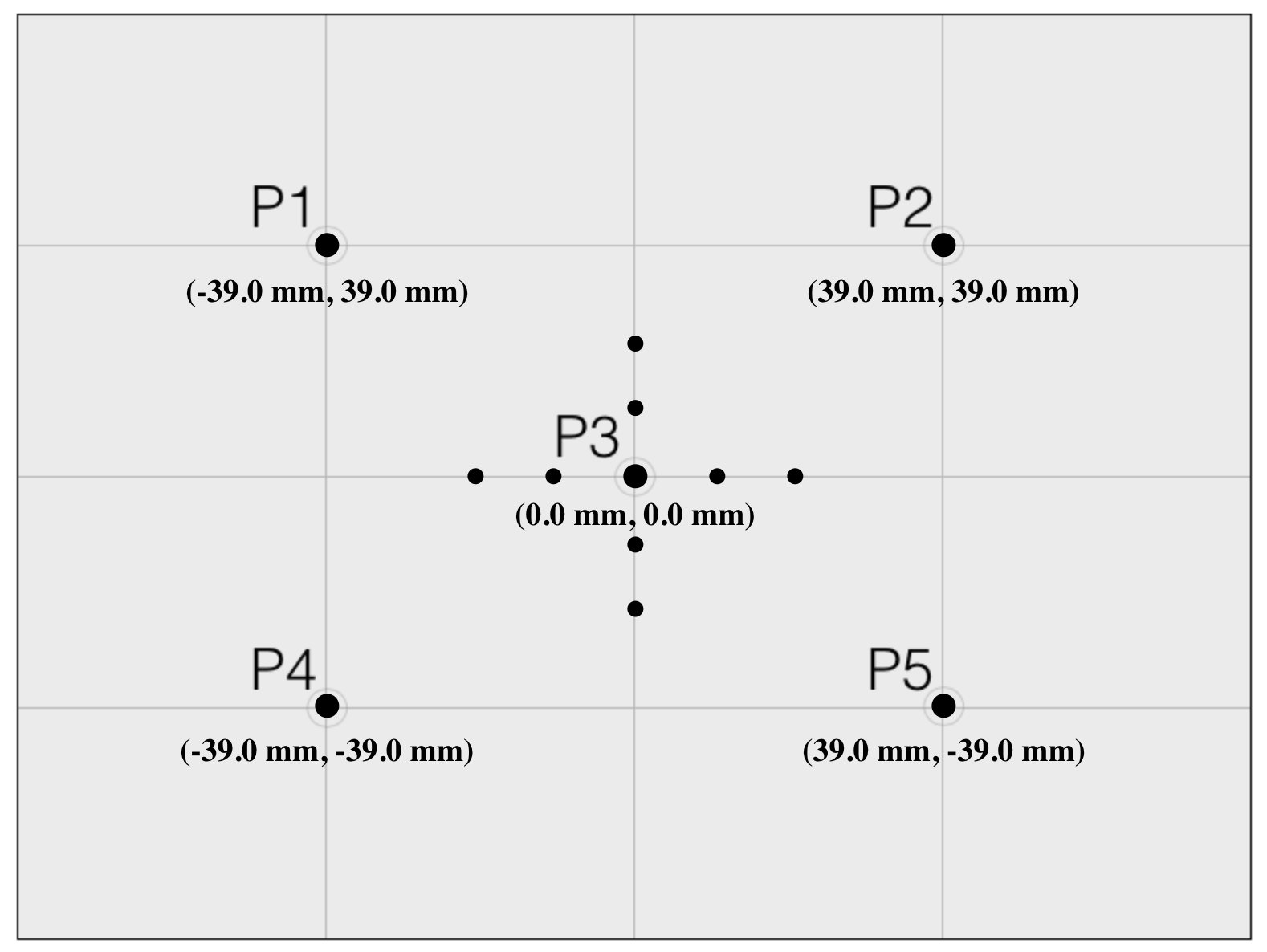}
      \caption{Reference positions on the fixture for a quadrant }
       \label{fig:czti_refpositions} 
\end{figure}

The radioactive source was positioned at different locations on the quadrant using the fixture.
Five reference positions P1, P2, P3, P4, and P5, directly above the intersection of four adjacent detector modules, were selected to place the source. Figure~\ref{fig:czti_refpositions} shows the five selected positions. 

\begin{figure}
   \centering
     \includegraphics[width=0.5\textwidth]{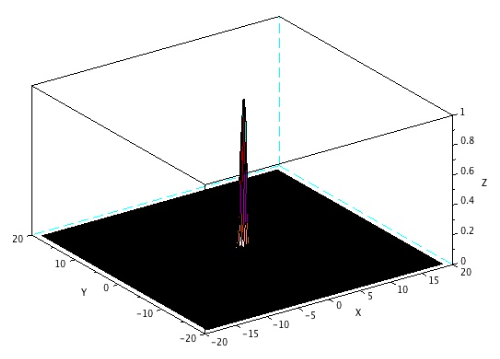}
        \caption{Reconstructed source profile using the Richardson-Lucy method when the source is placed at the central reference P3 during FM ground calibration.}
       \label{fig:ground_cal} 
\end{figure}

In addition to the five reference positions, the radioactive source was also placed at a few locations, shifted in the X and Y directions from the central reference P3, as shown in Fig.~\ref{fig:czti_refpositions}. The height of the source plane was 2~metres above the detector. The uncertainty in the horizontal and vertical positioning of the source is estimated to be $\sim 2.5$~mm and $\sim 1$~cm respectively. The source illumination axis was kept perpendicular to the detector within an accuracy of about 0.14$\degree$. A library of shadows was created by computing expected shadow patterns for a range of source positions, spanning 39~mm on either side of each reference position in both directions in steps of 0.25~mm. The acquired datasets were then analyzed using cross-correlation and Richardson-Lucy algorithms. Figure~\ref{fig:ground_cal} shows the reconstructed source profile using the Richardson-Lucy algorithm for an observation where the radioactive source is located at position P3.

Table~\ref{tab_fm_results} shows the deviation, in mm, of the reconstructed locations of the sources from the expected ones.
In the flight model calibration, the source was at the height of 2000~$mm$, whereas during the qualification model calibration source was at 897~$mm$ above the detector. Hence, in the flight model calibration, we see more deviation than the deviation seen the qualification model calibration.

\begin{table*}[ht!]
	\centering
    \begin{tabular}{|c|c|c|c|c|c|}
        \hline
	    Sr.No. & Quadrant ID & Lab Location & Source & CC (in mm) &RL (in mm)\\
        \hline
1   &     0   &    P1 &    Am     &(3.00, 3.00)     &(2.50, 1.50) \\ \hline
2   &     0   &    P3 &    Am &    (3.00, 2.25) &    (2.75, 2.00)\\ \hline
3   &   0   &    P4 &    Am &    (-0.25, 1.75) &    (-0.50, 1.75)\\ \hline
4   &    0   &    P5 &    Am &    (-2.00, 4.00) &    (-2.00, 4.00)\\ \hline
5   &    0   &    P3 &    Co &    (-2.00, 2.00) &    (-1.75, 1.50)\\ \hline
6   &    0   &    P3+(-10,0) &Co &     (0.50, 2.00) &    (-0.50, 1.50)\\ \hline
7   &    0   &    P3+(-20,0) &Co &    (0.00, 2.25) &    (-0.50, 2.00)\\ \hline
8   &      0   &    P3+(10,0)  &Co &    (0.75, 2.00) &    (0.75, 2.00)\\ \hline
9  &    0   &    P3+(20,0)  &Co &    (1.25, 2.25) &    (1.00, 2.00)\\ \hline
10  &    0   &    P3+(0,10)  &Co &    (0.50, 1.00) &    (0.50, 1.00)\\ \hline
11  &    0   &    P3+(0,20)  &Co &    (2.00, -1.75) &    (1.00, -1.75)\\ \hline
12  &     0   &    P3+(0,-10) &Co &    (2.00, -3.00) &    (0.00, -3.75)\\ \hline
13  &    0  &    P3+(0,-20) &Co &    (-1.25, -2.75) &(-2.00,- 2.75)\\ \hline

14  &    0  &    P2     &Co     &(-1.25, -2.00)     &(-1.25, -1.75)\\ \hline
15  &    0  &    P4     &Co     &(-3.00, 1.75) &    (-2.50, 1.75)\\ \hline
16  &    0  &    P5     &Co     &(-2.25,1.50) &    (1.25 ,1.75)\\ \hline

17 &   1   &    P3	    &Co & (0.25,-0.50)    &    (0.50, -0.25)\\ \hline
18 &   1   &    P3+(-10,0) &Co & (0.50, 1.25)    &    (0.00, 1.00)\\ \hline
19 &   1   &    P3+(10,0)  &Co & (0.00, 0.50)    &    (0.25, 0.50)\\ \hline
20 &   1   &    P3+(0,10)  &Co & (0.25,0.25)     &    (0.25, 0.0)\\ \hline
21 &   1   &    P3+(0,20)  &Co & (0.75,-0.50)    &    (0.50,-0.25)\\ \hline
22 &   1   &    P3+(0,-10) &Co & (0.00,-1.25)    &    (0.25, -1.00)\\ \hline
23 &   1   &    P3+(0,-20) &Co & (1.0, -1.00)    &    (0.75, -1.25)\\ \hline

24 &    2  &    P3     &Co     &(1.25, 0.75) &    (2.25, 0.75)\\ \hline
25 &    2  &    P3+(-10,0) &Co &    (1.00, 0.75) &    (0.75, 0.75)\\ \hline
26 &    2  &    P3+(-20,0) &Co &    (1.00, 1.00) &    (0.00, 1.25)\\ \hline
27 &    3  &    P1     &Am     &(-0.75, 3.00) &    (-1.25, 1.00)\\ \hline
28 &    3  &    P2     &Am     &(3.00, 2.25) &    (3.25, 2.25)\\ \hline
29 &    3  &    P3     &Am     &(0.75, 0.75) &    (0.75, 0.75)\\ \hline
30 &    3  &    P4     &Am     &(2.25, 1.25) &    (2.50, 1.00)\\ \hline

31 &    3  &    P3     &Co     &(0.00, 1.25) &    (1.25, 1.00)\\ \hline
32 &    3  &    P3+(-10,0) &Co &    (-1.00, 0.00) &    (-1.00, 0.00)\\ \hline
33 &    3  &    P3+(-20,0) &Co &    (-0.50, 0.00) &    (-1.50, 0.25)\\ \hline
34 &    3  &    P3+(10,0)  &Co &    (0.75, 0.00) &    (1.00, 0.75)\\ \hline
35 &    3  &    P3+(20,0)  &Co &    (1.00, 0.00) &    (1.75, 1.25)\\ \hline
36 &    3  &    P3+(0,10)  &Co &    (0.25, -0.75)&    (0.25, -0.75)\\ \hline
37 &    3  &    P3+(0,20)  &Co &    (1.50, -0.50) &    (1.50,0.50)\\ \hline
38 &    3  &    P3+(0,-10) &Co &    (0.25, -1.00) &    (0.25, -1.00)\\ \hline
39 &    3  &    P3+(0,-20) &Co &    (1.25, -2.00) &    (1.00, -2.00)\\ \hline
40 &    3  &    P1     &Co &    (-1.00, 2.00) &    (-2.00, -0.50)\\ \hline
41 &    3   &    P2     &Co &    (2.25, 2.25) &    (0.00, 2.25)\\ \hline
42 &    3   &    P4     &Co &    (0.00, 0.00) &    (0.00, 0.00)\\ \hline
43 &    3   &    P5     &Co &    (1.00, 0.75) &    (1.00, 0.75)\\ \hline
    \end{tabular}
    \caption[FM ground calibration reconstruction results ]{ Laboratory source positions set during FM ground calibration and the offsets in reconstructed source positions by  Cross-correlation and Richardson-Lucy algorithms. Units are millimetres.}
    \label{tab_fm_results}
\end{table*}

\begin{figure}
   \centering
     \includegraphics[width=0.5\textwidth]{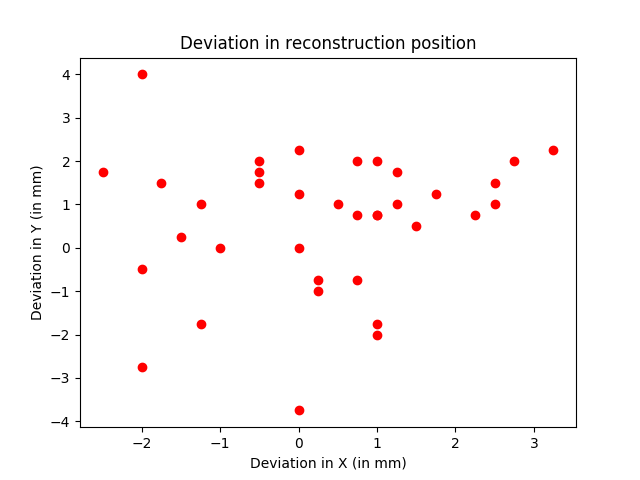}
    \caption[Deviations in the reconstructed positions in FM calibration ]{Deviation of the source positions recovered using the Richardson-Lucy algorithm with respect to the expected values during the FM ground calibration.}
       \label{fig_fm_deviations} 
\end{figure}

Figure~\ref{fig_fm_deviations} shows a scatter plot of deviations in the reconstructed locations from those expected. In most cases, the deviation is well within 1.5~mm except for a small fraction extending up to 3~mm. The deviation of 3~mm at a height of $2000~mm$ corresponds to an angular deviation of 5~arcmin, within the design goal of 8~arcmin for the CZTI. 

\subsection{In-flight Calibration}\label{inflight_cal}
\begin{figure*}[ht!]
	
	\begin{subfigure}{0.45\textwidth}
	    \centering
		\includegraphics[width=\textwidth]{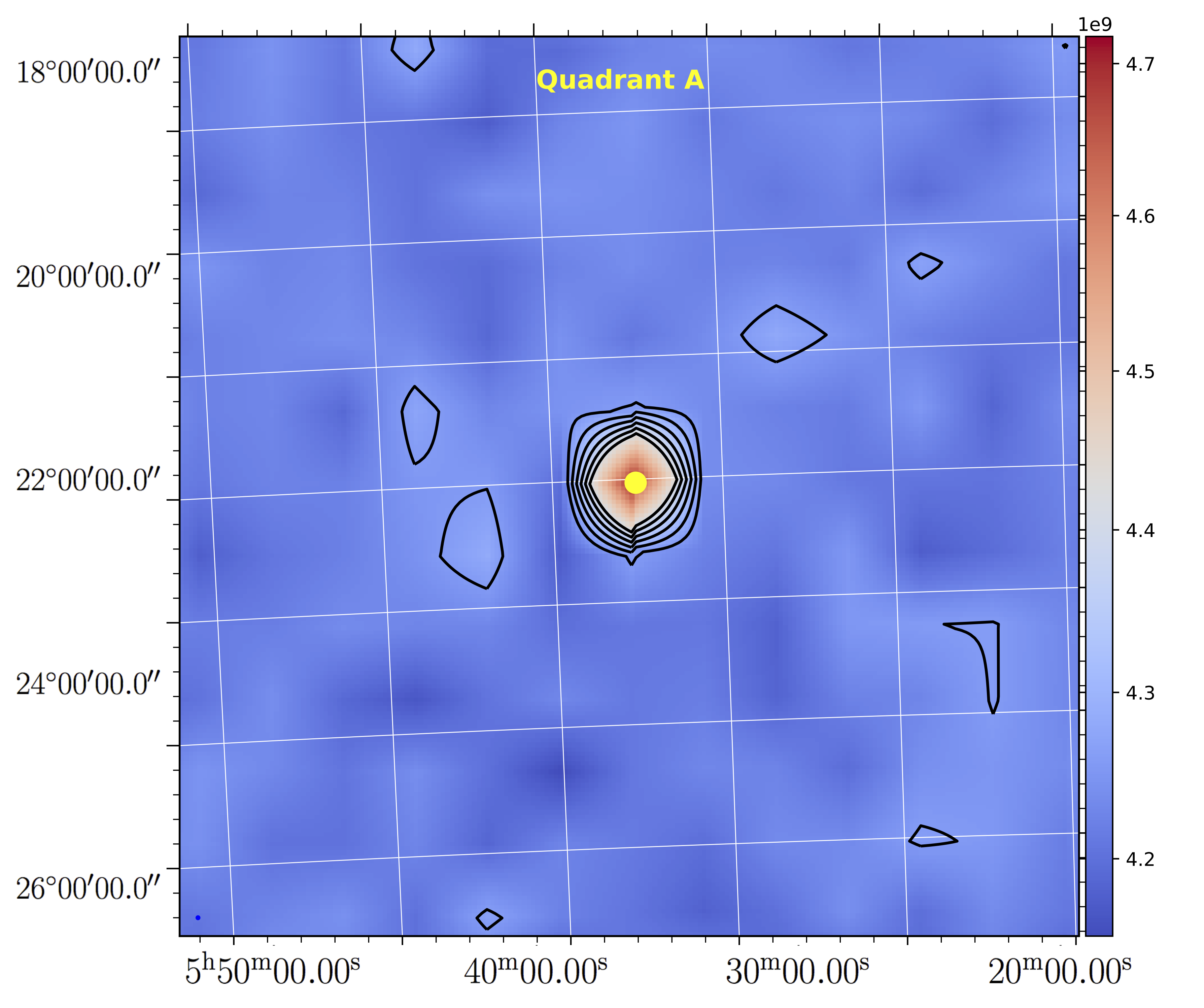}
		\caption{Quadrant A. }
	\end{subfigure}
	\begin{subfigure}{0.45\textwidth}
	    \centering
		\includegraphics[width=\textwidth]{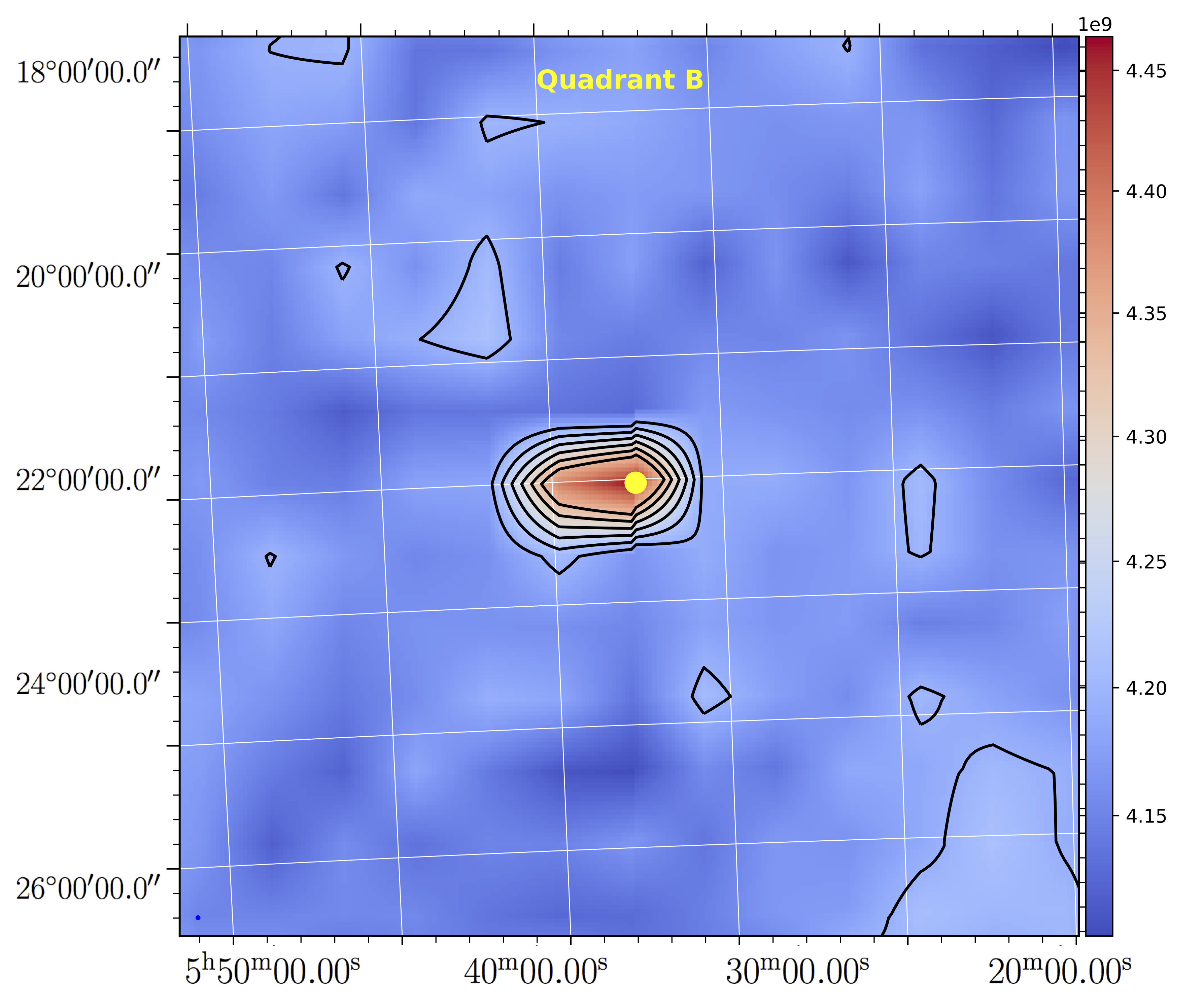}
		\caption{ Quadrant B.}
	\end{subfigure}
	\newline
	\begin{subfigure}{0.45\textwidth}
	    \centering
		\includegraphics[width=\textwidth]{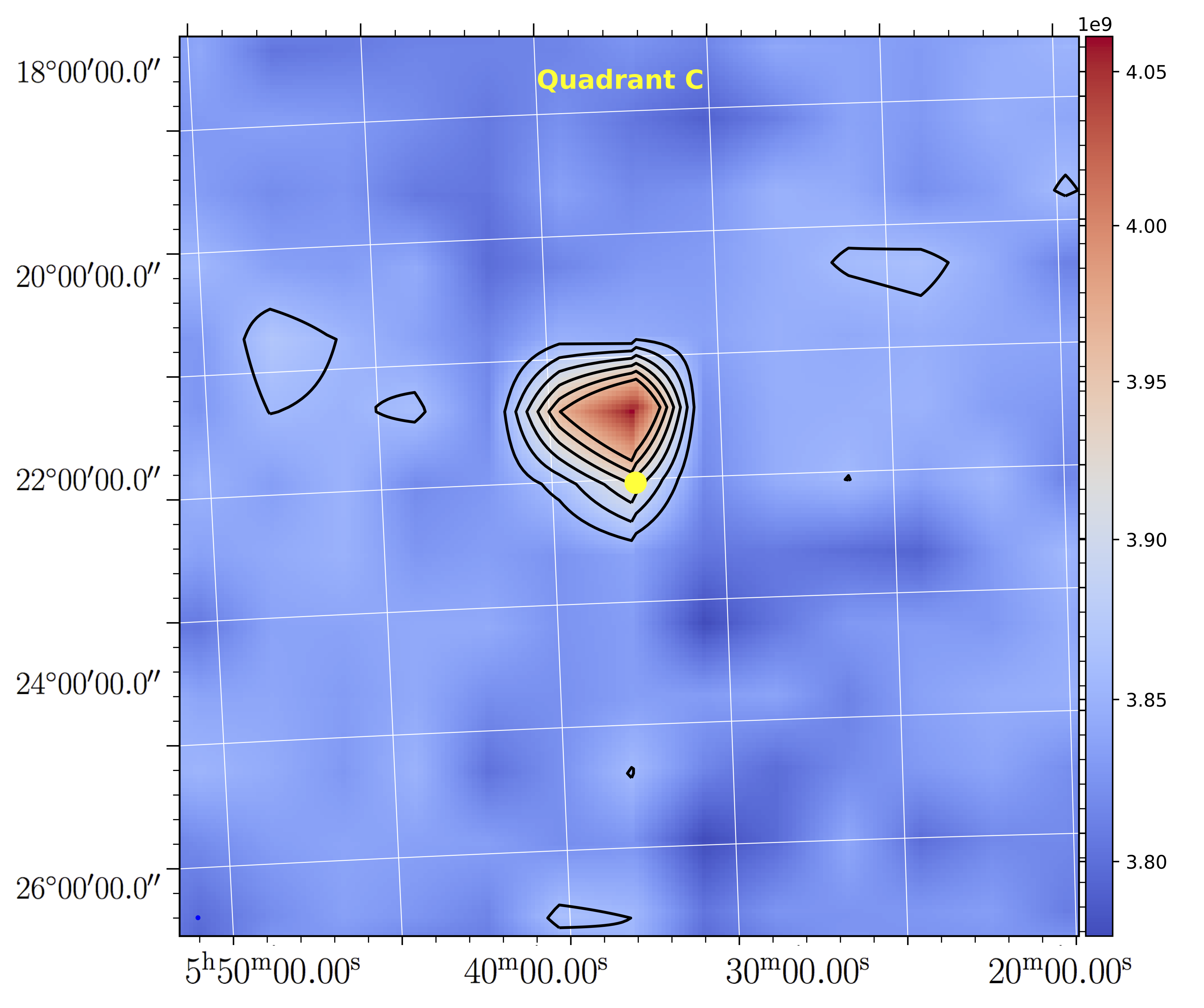} 
		\caption{Quadrant C.}
	\end{subfigure}
	\begin{subfigure}{0.45\textwidth}
	    \centering
		\includegraphics[width=\textwidth]{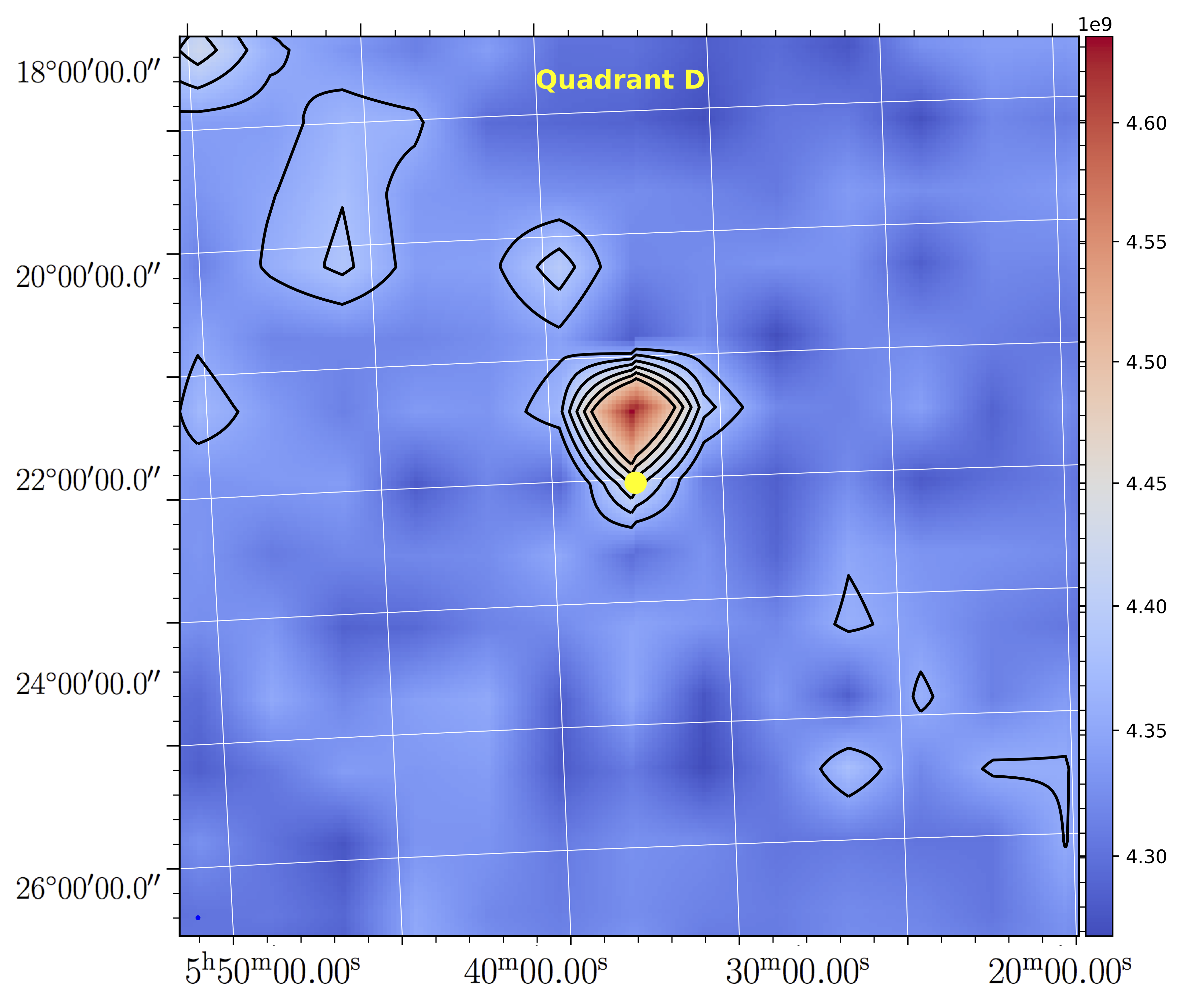} 
		\caption{Quadrant D.}
	\end{subfigure}
	\caption{Image of the Crab Nebula reconstructed from CZTI data using mask cross-correlation technique for observation id 9000000406.  Note that the angular extent of the source is well within the resolution of the CZTI, so the image profile primarily reflects the angular response of CZTI.}
	\label{fig:fft_image}
\end{figure*}

The first six months after the launch of AstroSat was reserved for the Performance Verification (PV) and calibration of all the instruments. During this period, several calibration observations were conducted of the Crab nebula, one of which, observation ID 20160331\_T01\_112T01\_9000000406, was used to characterise the post-launch imaging performance.  This observation was performed from UT 2016-03-31 05:39:19 to 2016-04-03 03:55:11, adding up to a total of 114~ks exposure time. The observed data was analyzed using the CZTI data analysis pipeline\footnote{http://astrosat-ssc.iucaa.in/?q=cztiData}. Figure~\ref{fig:fft_image} shows the reconstructed sky image of the Crab from all four quadrants of CZTI using the mask cross-correlation technique. In the images, the expected location (shown as a yellow circle) of the source matches the peak of the reconstructed profile for quadrants A and B, while images from quadrants C and D show a deviation of 0.29$\degree$ in the Y direction. For quadrant B, although the reconstructed source location is correct, there is an extended tail in the X-direction. 

To understand the discrepancy between the expected and the reconstructed images, various experiments and simulations were performed. For quadrant A, the reconstruction was as expected, so the experiment started with quadrant B, which showed an extended peak. 
The following possible scenarios were investigated to understand the cause of the extension and the shifts observed in the reconstructed image.

\begin{enumerate}
\item Extension or shift because of measurement error in the distance between the detector and the mask
\item Extension or shift because of unstable pointing
\item Extension or shift because of module dependency
\item Extension or shift because of misalignment between the detector and mask
\end{enumerate}

\subsubsection{Extension or shift because of measurement error in the distance between the detector and the mask}
\begin{figure*}[ht!]
    \centering
	\begin{subfigure}{0.45\textwidth}
	    \centering
		\includegraphics[width=\textwidth]{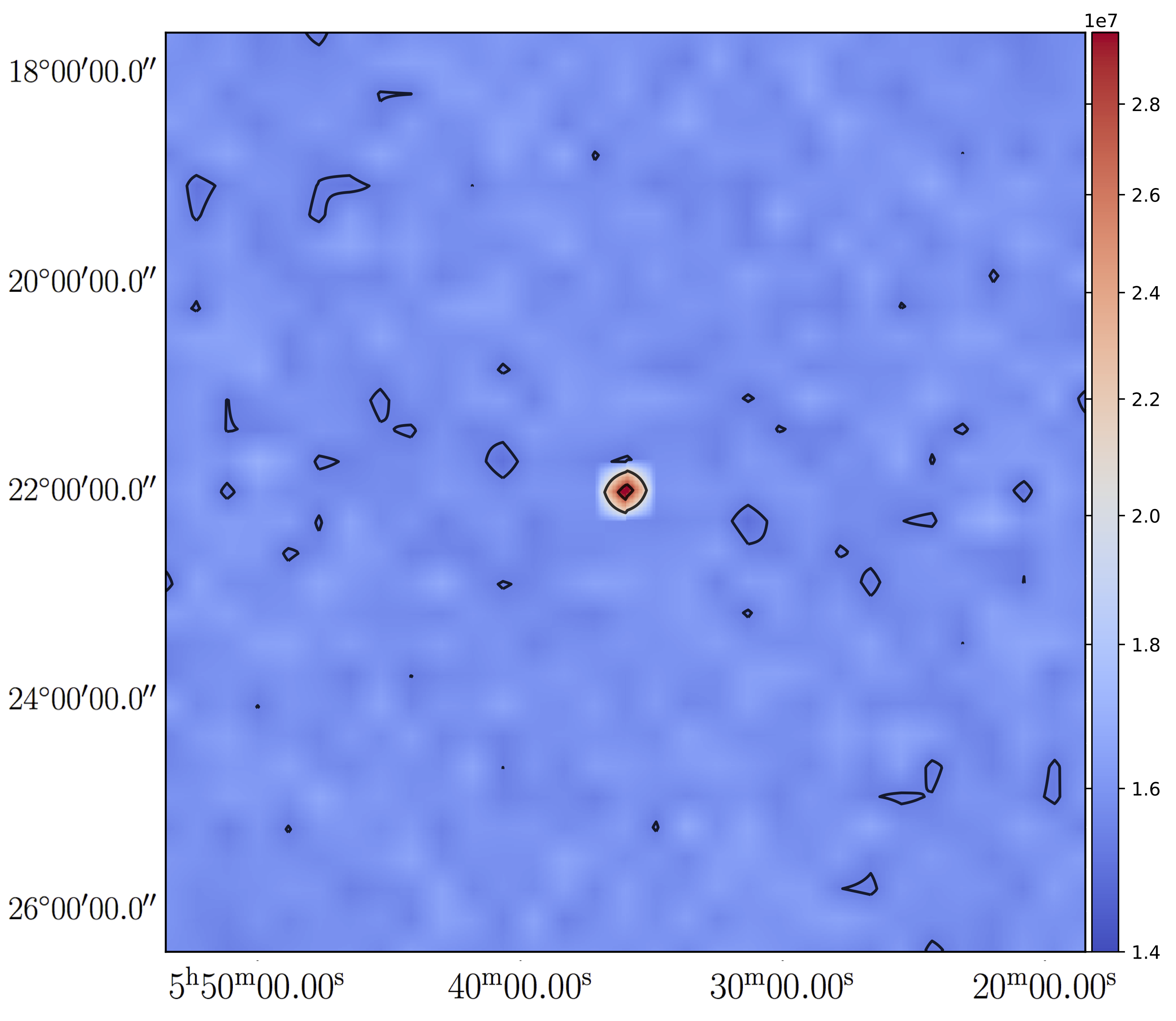}
		\caption{Mask plate height $400~mm$ from detector.}
	\end{subfigure}
	\begin{subfigure}{0.45\textwidth}
	    \centering
		\includegraphics[width=\textwidth]{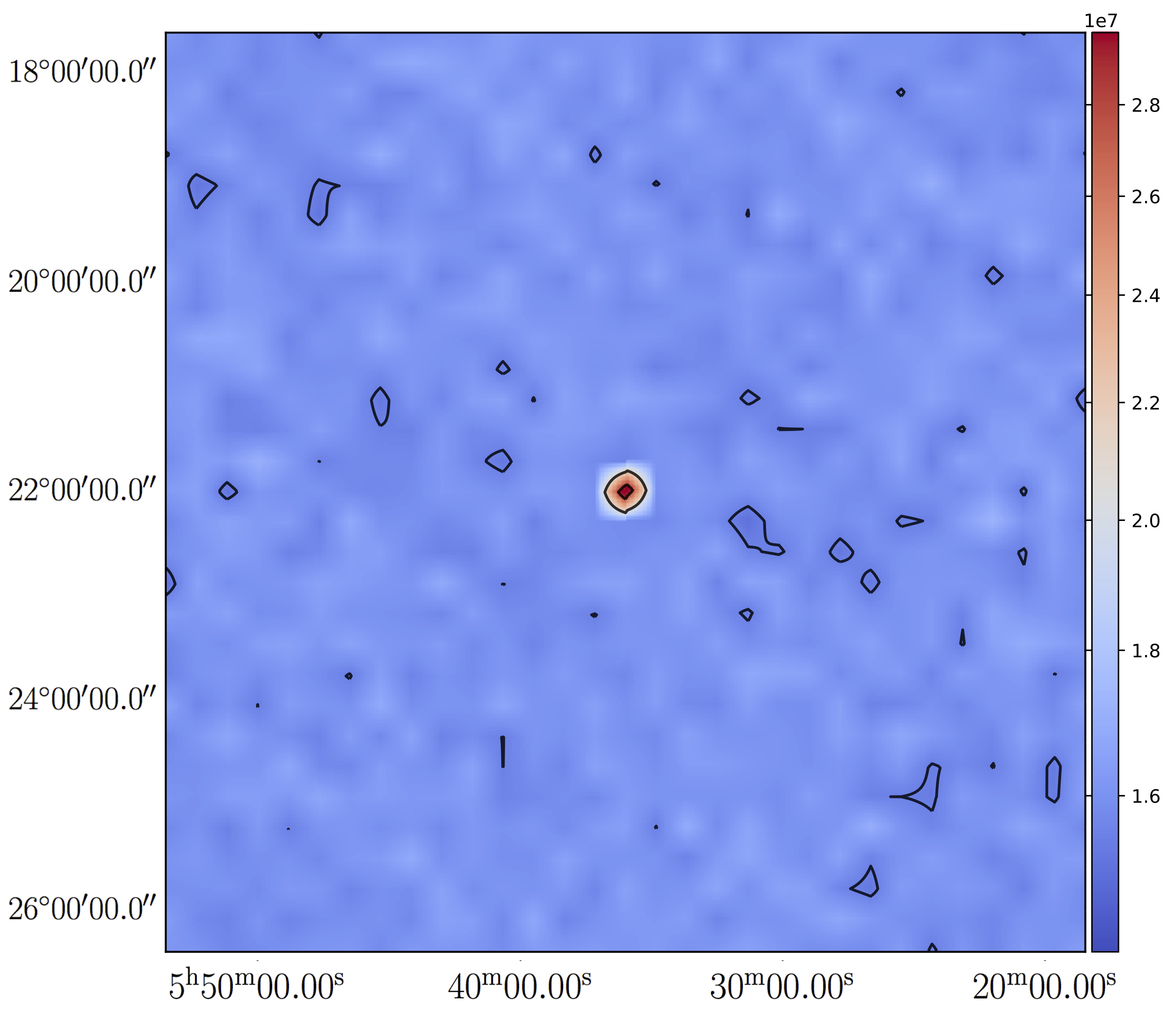}
		\caption{Mask plate height $450~mm$ from detector.}
	\end{subfigure}
	\caption{Image reconstructed from simulated DPI with mask plate placed at a height $400~mm$ (Left panel) and $450~mm$ (Right panel)}
	\label{fig_sim_maskheight}
\end{figure*}

The distance between the coded aperture and the detector is a key factor in the design of a coded mask instrument. A wrong value for this used in the reconstruction process could lead to distortions in the recovered image. In the case of CZTI, the mask plate is placed 481~mm above the detector as per design.  
To investigate the effects of a possible error in this value, we simulated DPIs with the mask placed at heights 470~mm, 460~mm, 450~mm and 400~mm above the detector, and subjected them to the same image reconstruction as employed for the real data.  This is a much exaggerated range of heights than the error margin that could be expected in reality. Figure~\ref{fig_sim_maskheight} shows the images reconstructed from simulated DPI with the mask plate placed at 450~mm and 400~mm respectively. The reconstructed images show neither an extension nor a shift. Hence, an error in mask height was ruled out as a possible cause of the image distortions observed.

\subsubsection{Extension or shift because of planar misalignment between the mask and detector}

\begin{figure*}[ht!]
    \centering
	\begin{subfigure}{0.45\textwidth}
		\includegraphics[width=\textwidth]{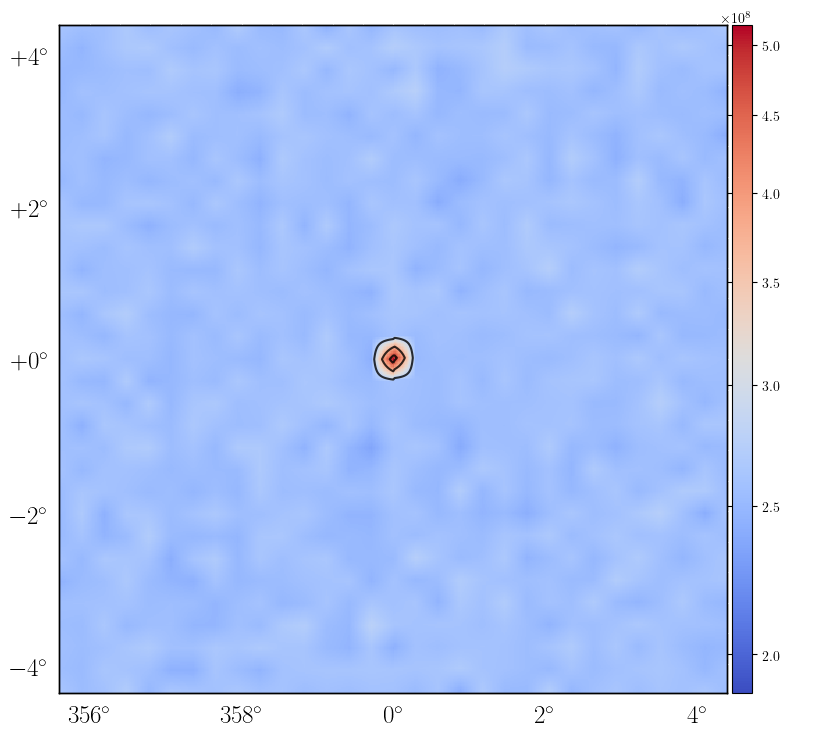}
		\caption{Mask plate tilted with 5 degree from the detector plane.}
	\end{subfigure}
	\begin{subfigure}{0.45\textwidth}
		\includegraphics[width=\textwidth]{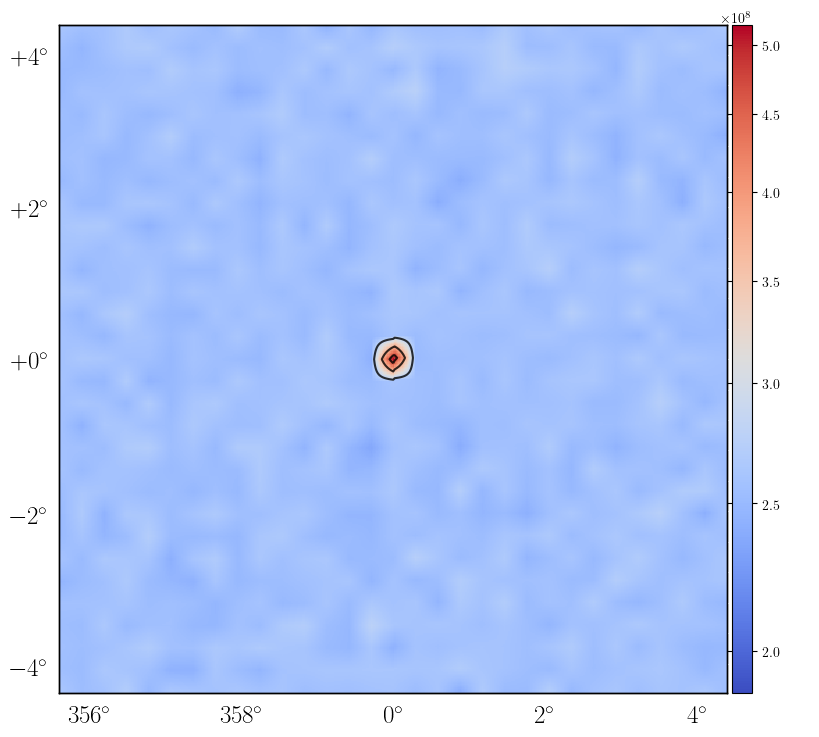}
		\caption{Mask plate tilted with 10 degree from the detector plane.}
	\end{subfigure}
	\caption{Reconstructed images by introducing planar misalignment between the mask and detector.}
	\label{fig_sim_planer}
\end{figure*}

The image reconstruction depends strongly on the planar alignment between the coded aperture and the detector, including any offset and rotations. If the mask and the detector are misaligned, then it can affect the reconstructed image. To examine the effect of a possible planar misalignment between the mask and detector, we simulated DPIs with planar misalignment of 1$\degree$, 3$\degree$, 5$\degree$,10$\degree$ and reduced these DPIs using the image reconstruction method. Figure~\ref{fig_sim_planer} shows the images reconstructed from simulated DPI with the mask plate tilted with 5$\degree$,10$\degree$. The reconstructed images show neither an extension nor a shift. Hence, we discarded the possibility of planar misalignment between the mask and the detector as the reason behind the image distortions observed.

\subsubsection{Extension or shift because of unstable pointing}
\begin{figure*}[ht!]
	\centering
	\begin{subfigure}{0.45\textwidth}
		\includegraphics[width=\textwidth]{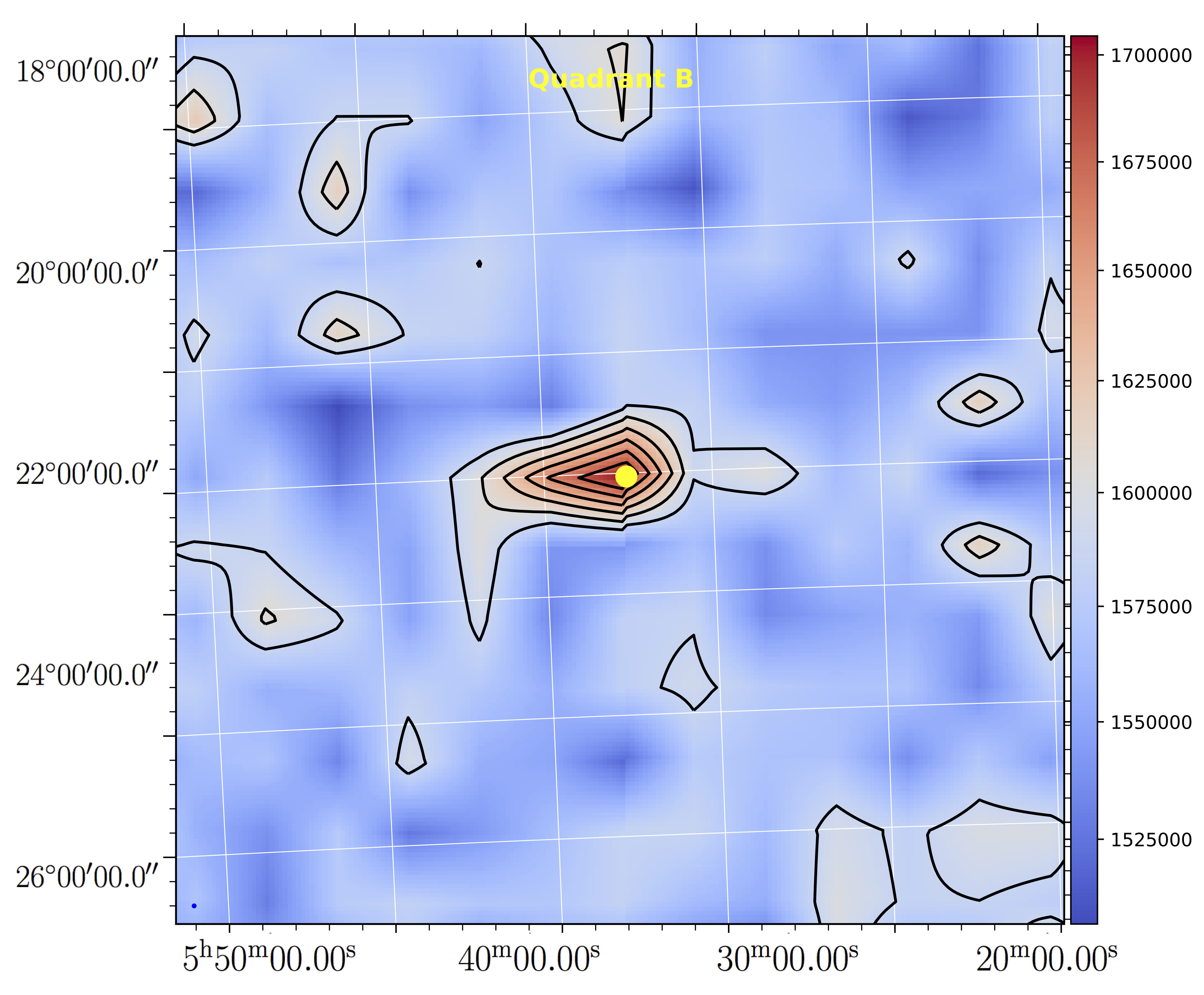}
	\end{subfigure}
	\begin{subfigure}{0.45\textwidth}
		\includegraphics[width=\textwidth]{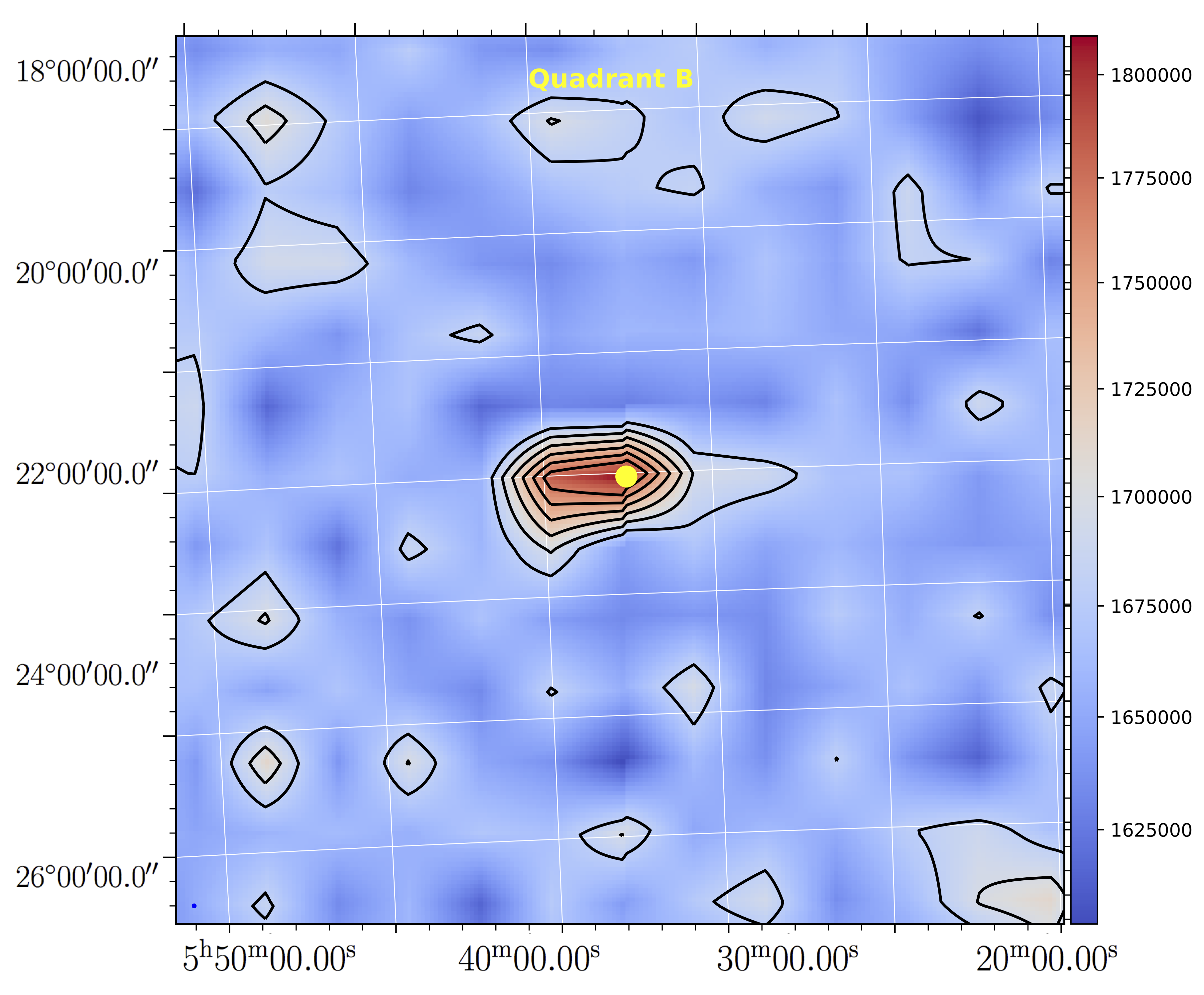}

	\end{subfigure}
	\caption{Reconstructed image at two time bins for Quadrant B}
	\label{fig_timebin}
\end{figure*}

If the time span over which the data for the DPI ere accumulated saw a significant variation in the pointing of the CZTI, then the reconstructed image could show a smearing.  Individual Crab images can be constructed from the CZTI data for integration times as short as 30~sec. Doing so with quadrant B data showed  very similar extension in individual images (Figure~\ref{fig_timebin}).  Examination of the satellite attitude data, collected at 128~ms sampling, revealed that the attitude jitter was no more than 0.045 degree rms over the time scale of minutes to hours, thus making it very unlikely that the observed image extension could result from it.

\subsubsection{Module dependence}
\begin{figure}
    \centering
    \includegraphics[width=0.5\textwidth]{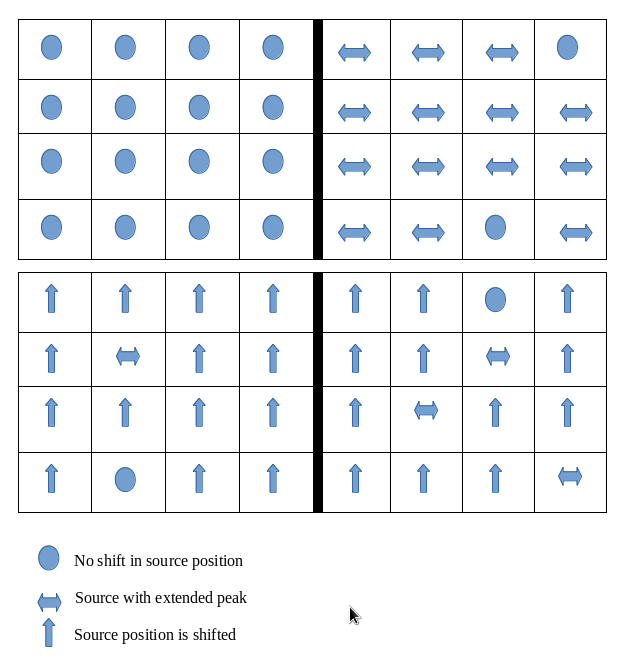}
    \caption{Distortions observed in module level imaging}
    \label{fig:module_image}
\end{figure}

Each CZTI quadrant is composed of 16 detector modules, all of which operate independently. At this point it was essential to establish whether the image distortion observed was a quadrant level phenomenon or do individual detector modules in a quadrant display discordant behaviour. We therefore reconstructed images using events from individual modules in all four quadrants.  In the reconstructed images, no module in the quadrant A showed any shift or extension. However, in quadrant B, 14 modules showed extension and in quadrants C and D, 14 and 12 modules respectively showed the shift in the reconstructed source position. An account of this is presented schematically in  Figure~\ref{fig:module_image}.  This collective display strongly indicated that the image extension or shift was associated with the imperfection of the complete assembly at the level of quadrants.

\subsubsection{Detector-mask misalignment}
Results from module-level imaging lead us to the possibility of misalignment between the detector and the mask. We performed various simulations to examine the effect of misalignment on the reconstructed images.
Initially, we simulated a DPI by shifting the mask pattern by $0.5~mm$ in the negative X direction.  The source image reconstructed from the simulated DPI showed an extended peak.  
Next, we shifted the mask pattern at steps of $0.1$~mm in the negative X-direction and compared the simulated image profiles with the observed image profile.

\begin{figure*}[ht!]
	\centering
	\begin{subfigure}{0.45\textwidth}
		\includegraphics[width=\textwidth]{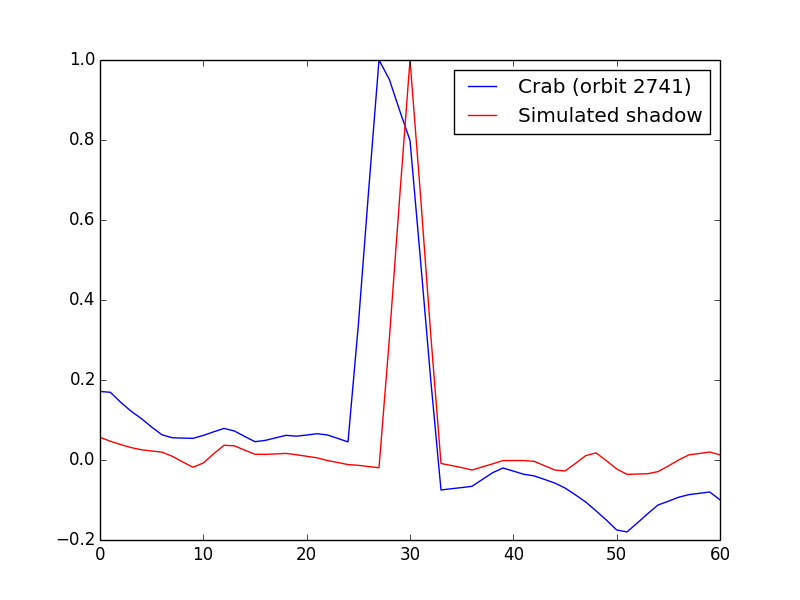}
		\caption{Without mask shift}
	\end{subfigure}
	\begin{subfigure}{0.45\textwidth}
		\includegraphics[width=\textwidth]{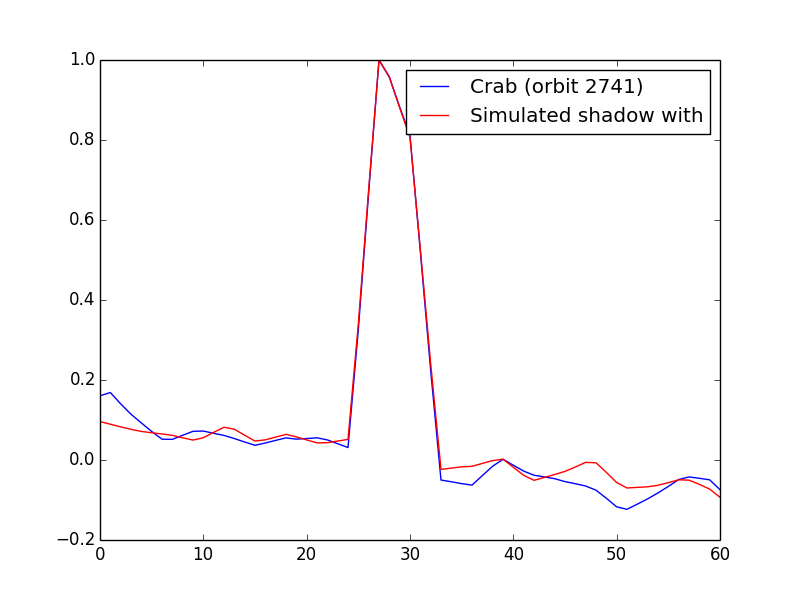}
		\caption{With mask shift}

	\end{subfigure}
	\caption{Simulated 1-d image profiles with zero (left panel) and $-0.5$~mm (right panel) shift of the mask pattern in the x-direction, compared with that obtained from observed data}
	\label{fig:q1_image_profiles}
	
\end{figure*}

\begin{figure}[ht!]
    \centering
    \includegraphics[width=0.5\textwidth]{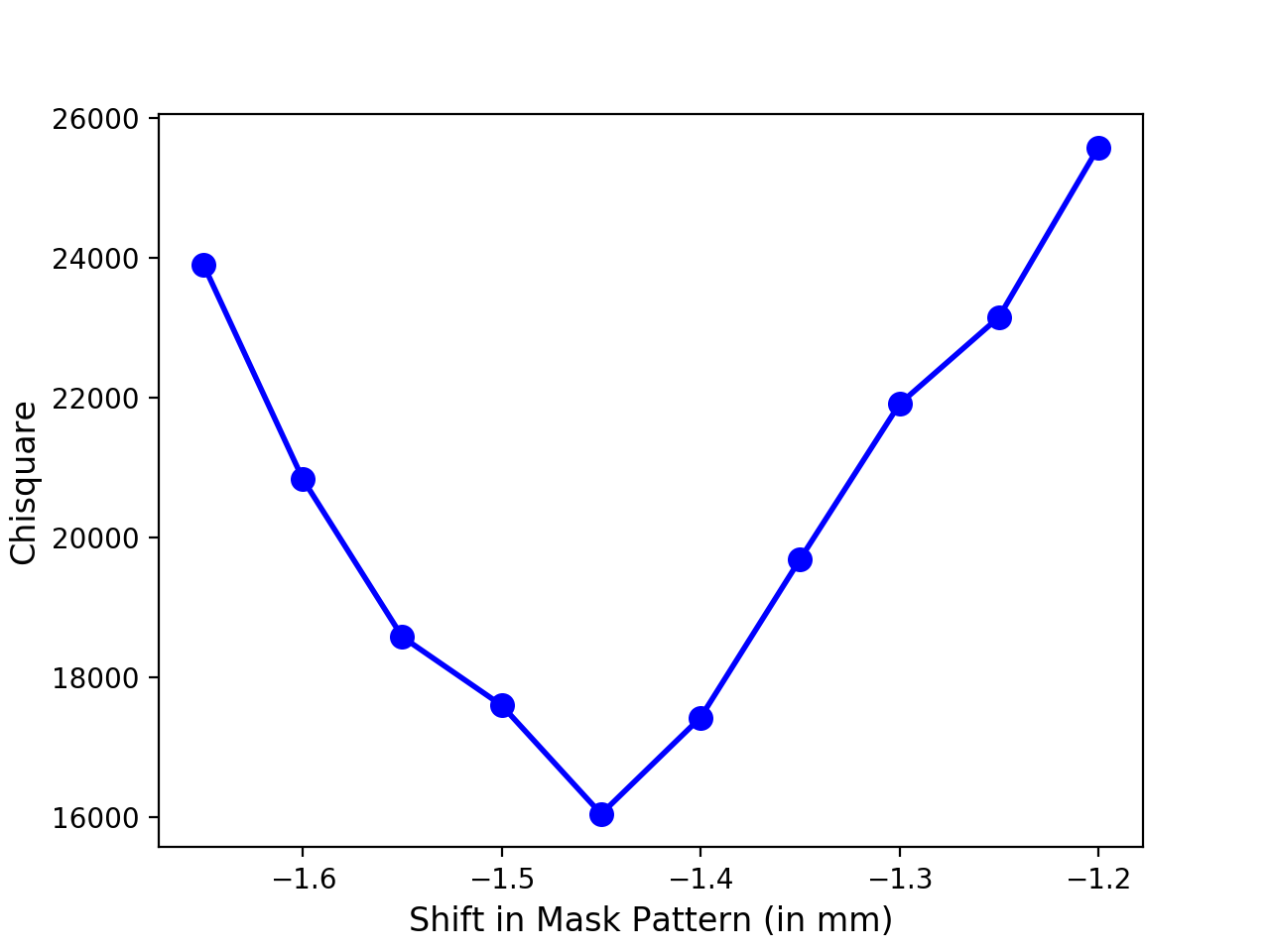}
    \caption[Chi-square value as a function of shift in the X direction for Quadrant B]{Chi-square value as a function of mask shift in the X direction for Quadrant B.}
    \label{fig_chisq}
\end{figure}

The one-dimensional image profile of the image reconstructed from observed data, and simulated data without any mask shift showed a strong disagreement. However, the image profile of the image reconstructed from simulated data with a mask shift of $-1.45$~mm showed a strong agreement with the observed image profile. 
Figure~\ref{fig:q1_image_profiles} shows the one-dimensional image profiles for observed and simulated data generated from the direct cross-correlation image.
Then, we simulated DPIs for shifts in the X-direction, from $-1.20$~mm to $-1.65$~mm with a step of  $0.05$~mm and compared the resulting images with that observed. The minimum chi-square was found for a mask shift of $-1.45$~mm, figure~\ref{fig_chisq} .  Hence, this experiment confirmed that in quadrant B there is a slight misalignment of $-1.45$~mm between the detector and the mask plate in the X-direction. Shifts for other quadrants were also estimated using a similar methodology. Table~\ref{tab_shift_results} below outlines the shift in the mask pattern with respect to the detector for all four quadrants.

\begin{figure}[ht!]
    	\centering
    	\includegraphics[width=0.5\textwidth]{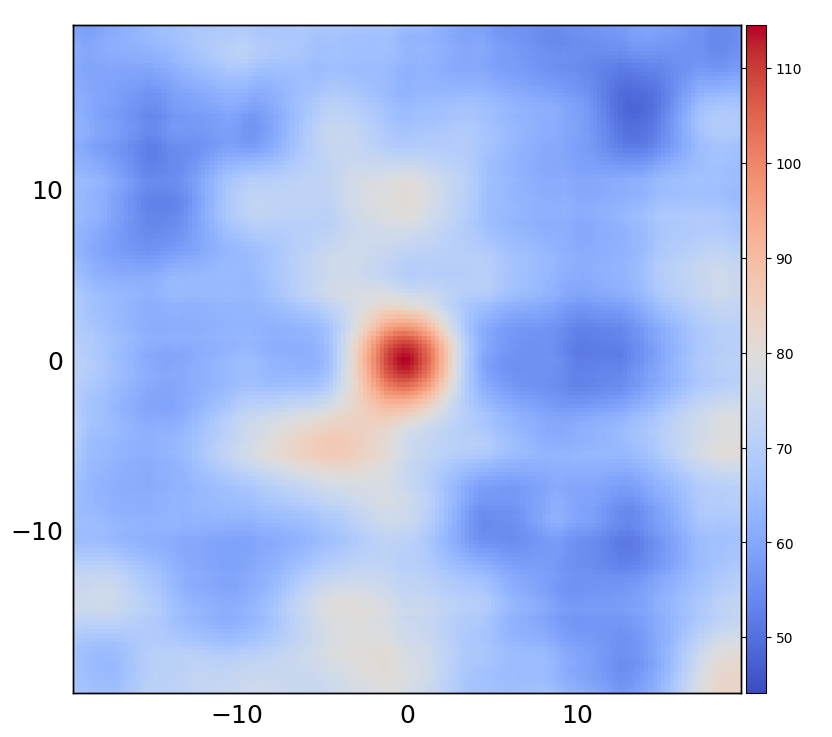}
	\caption{The image reconstructed from simulated DPI with the shifted mask in near field imaging similar to CZTI ground calibration.}
    \label{fig_sim_qm_mask_shift}
\end{figure}

We have not observed any extension in the reconstructed images during the on-ground calibration.  To review the effect of the mask shift on the reconstructed images in near field imaging, we have simulated DPIs for diverging beam radiations shined by a radioactive source under the circumstances similar to the ground calibration. We then subjected the simulated DPIs to the image reconstruction algorithm. Figure ~\ref{fig_sim_qm_mask_shift} shows the reconstructed image generated using the simulated DPH with the mask shift of -1.45$~mm$. The reconstructed images did not show any extension or shift in the source position.  In near field imaging, one mask element illuminates more than one detector pixel due to the diverging beam of radiations and possible that the mask shift may not affect the reconstructed image.

\begin{table}
\begin{center}
\begin{tabular}{|c|c|c|}
\hline
Quadrant & X shift (in mm)&Y shift (in mm)\\
\hline
A & 0.00 & 0.00\\
\hline
B & -1.45 & 0.00\\
\hline
C  & 0.00 & 1.68\\
\hline
D & 0.00 & 1.50\\
\hline
\end{tabular}
\end{center}
\caption[The shift in the mask pattern with respect to the detector for all quadrants.]{The shift in the mask pattern with respect to the detector for all quadrants.}
    \label{tab_shift_results}
\end{table}
After calculating the shifts for all quadrants, we corrected for shifts in the imaging algorithm. The section below describes the corrections applied.

\section{Correcting for Mask Shifts}
In mask cross-correlation, the DPI is cross-correlated with the coded aperture mask. To correct for the shift in the reconstructed image, we multiply a phase matrix in the Fourier domain, as shown in  Equation~\ref{eq:convphasemat}.

\begin{equation}
\mathcal{F}^{-1}(\mathcal{F}(DPI) X \mathcal{F}(MASK)*P)
\label{eq:convphasemat}
\end{equation}

Here, $\mathcal{F}$ is FFT, $\mathcal{F}^{-1}$ is inverse FFT, and P is a phase matrix as described in equation~\ref{eq:phasemat},

\begin{equation}
 P= e^{i 2\pi (f_x \delta_x +f_y \delta_y)}
\label{eq:phasemat}
\end{equation}

Here,\\
$\delta_x $ is the shift in X direction in pixel unit\\
$\delta_y $ is the shift in Y direction in pixel unit\\
$N_x$= number of pixels in X direction \\
$N_y$= number of pixels in Y direction \\
$f_x$= $\frac{-1}{N_x} +....+ \frac{1}{N_x}$\\
$f_y$= $\frac{-1}{N_y} +....+ \frac{1}{N_y}$\\

\begin{figure*}[ht!]
	\begin{subfigure}{0.45\textwidth}
	    \centering
		\includegraphics[width=\textwidth]{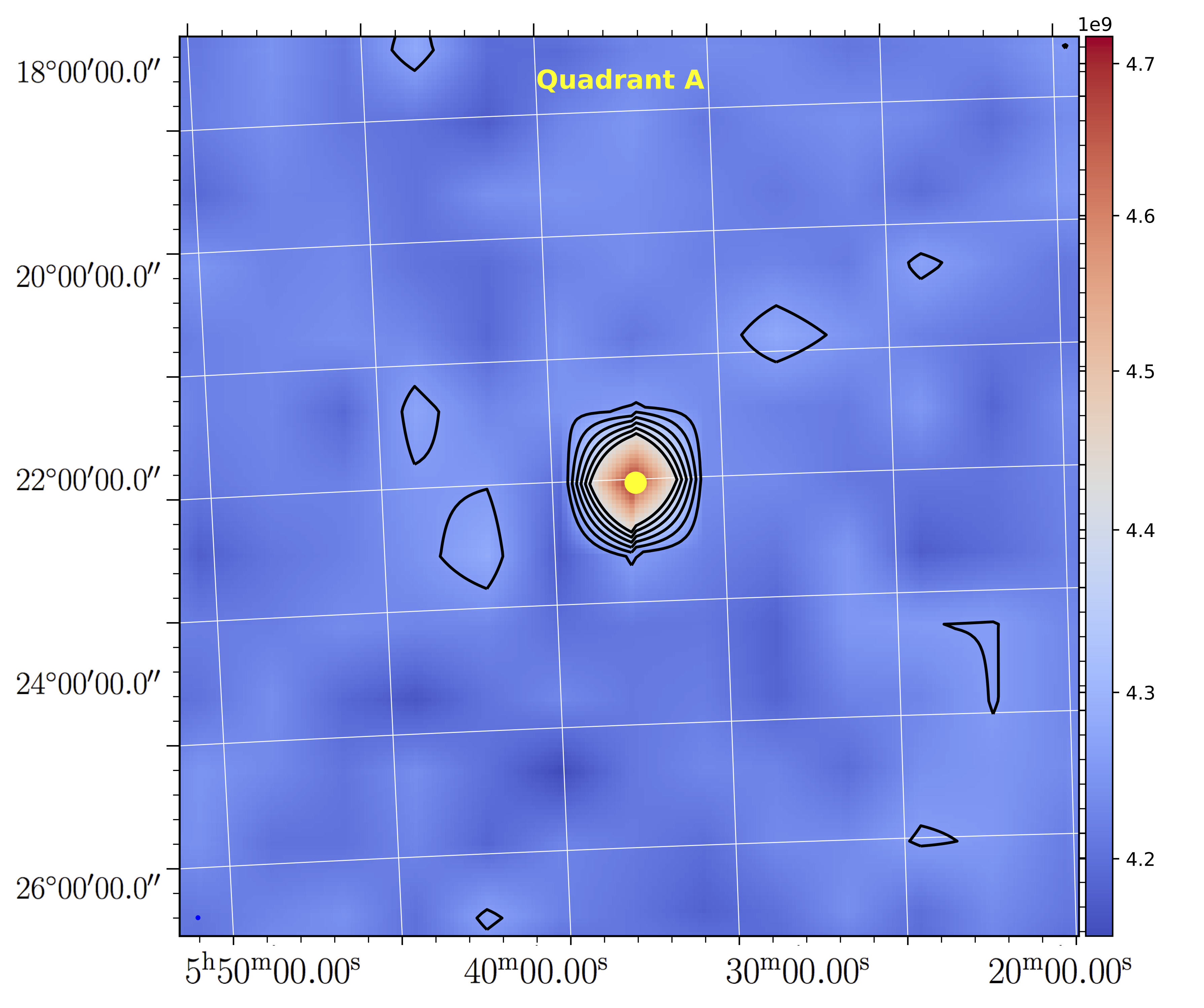}
		\caption{Quadrant A. }
	\end{subfigure}
	\begin{subfigure}{0.45\textwidth}
	    \centering
		\includegraphics[width=\textwidth]{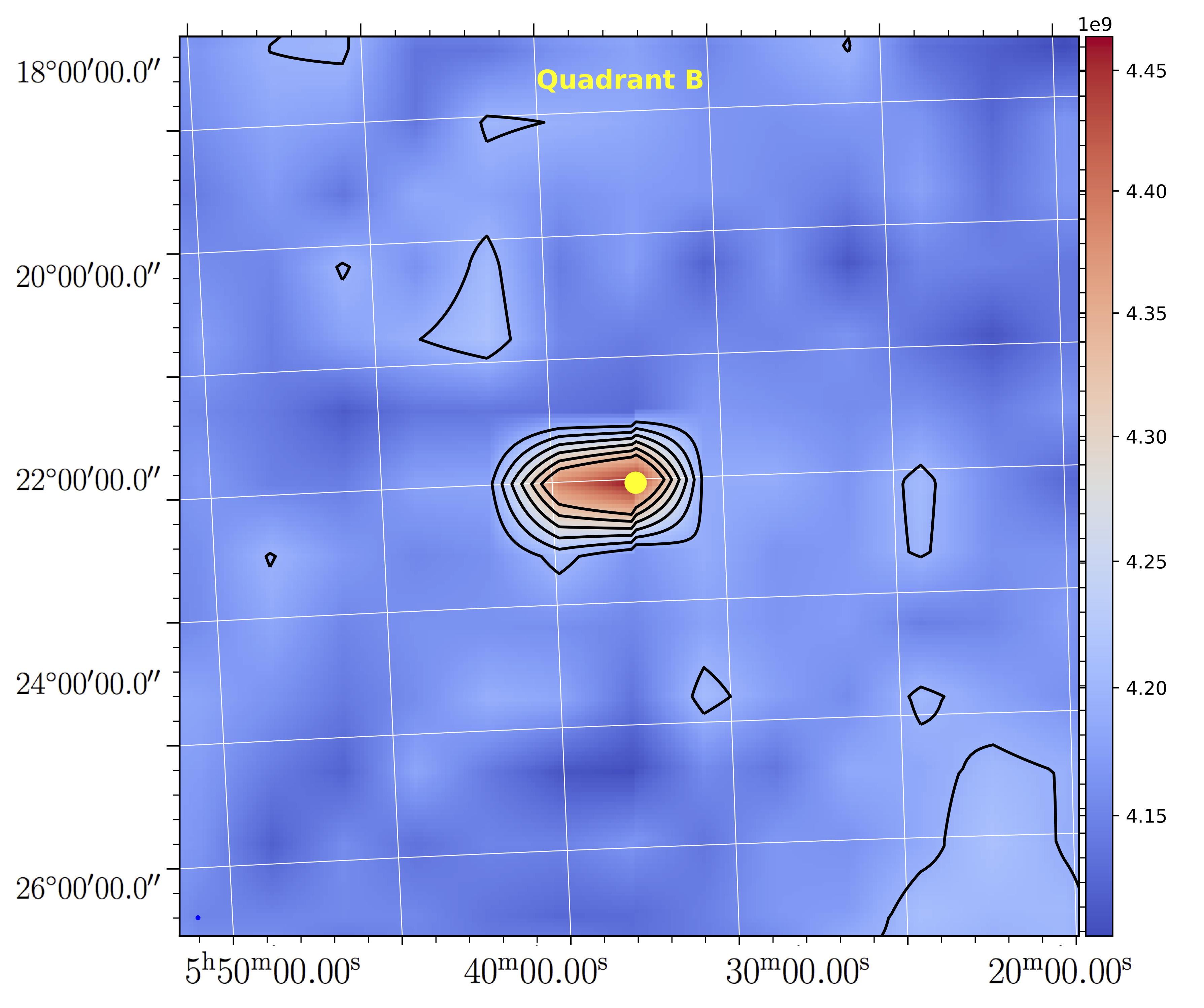}
		\caption{ Quadrant B.}
	\end{subfigure}
	\newline
	\begin{subfigure}{0.45\textwidth}
	    \centering
		\includegraphics[width=\textwidth]{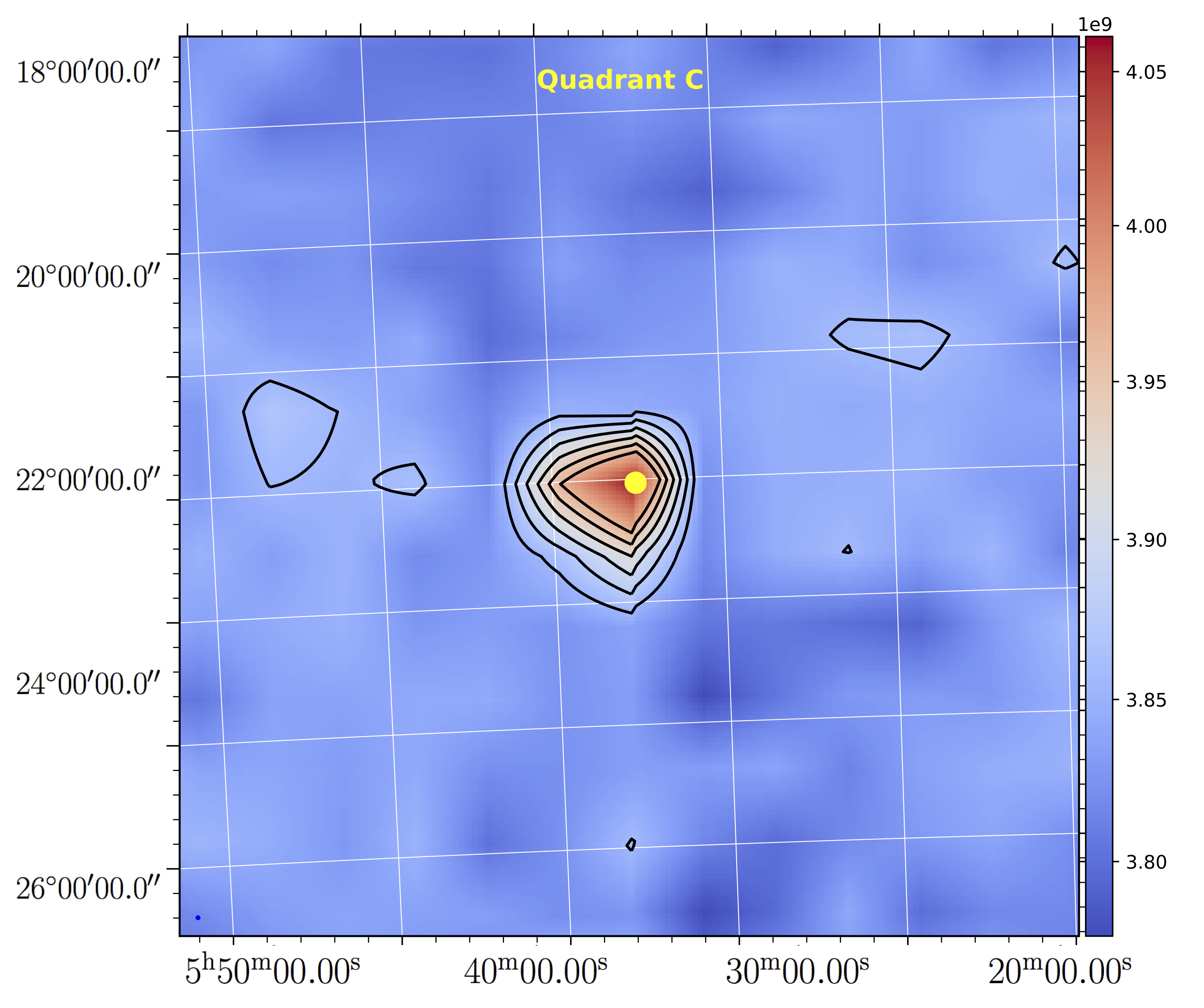} 
		\caption{Quadrant C. }
	\end{subfigure}
	\begin{subfigure}{0.45\textwidth}
	    \centering
		\includegraphics[width=\textwidth]{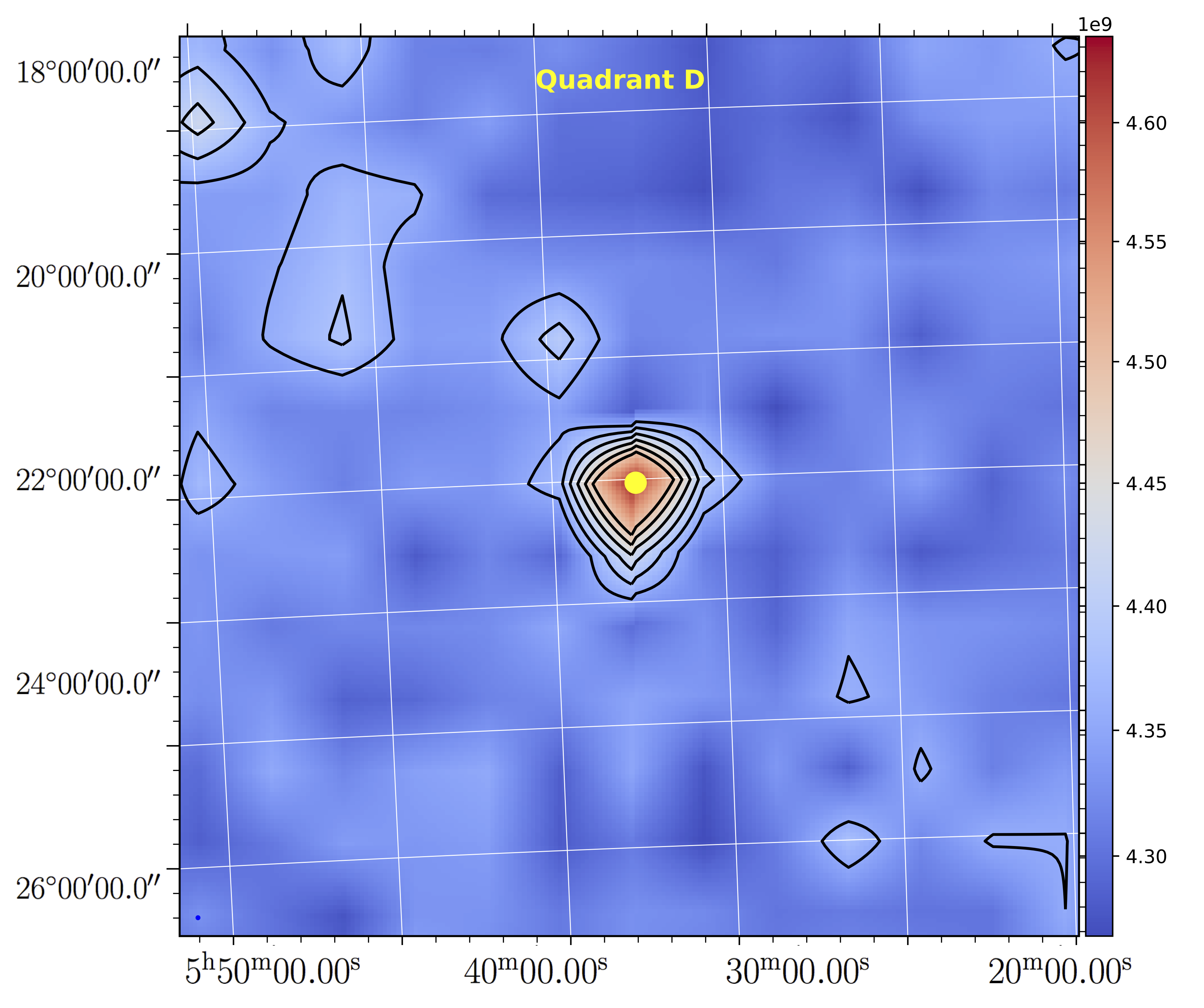} 
		\caption{Quadrant D.}
	\end{subfigure}
	\caption{Crab nebula image reconstructed using FFT after application of the phase matrix for observation id 9000000406.}
	\label{fig:fft_image_fixed}
	
\end{figure*}

After multiplication with the phase matrix and performing inverse FFT, the source is found to be at the expected location. CZTI data analysis pipeline is modified to incorporate the phase shift multiplication to correct the images reconstructed using mask cross-correlation images. Figure~\ref{fig:fft_image_fixed} shows the reconstructed sky images generated using modified mask cross-correlation for four quadrants. The phase matrix multiplication compensated for the shift in the reconstructed image, but the extension in quadrant B was not corrected. The extension in the image can be corrected using the shadow cross-correlation method (section~\ref{shadowcrosscorrelation}) using computed shadows that incorporate the measured mask shifts.

\begin{figure*}[ht!]
	
	\begin{subfigure}{0.45\textwidth}
	    \centering
		\includegraphics[width=\textwidth]{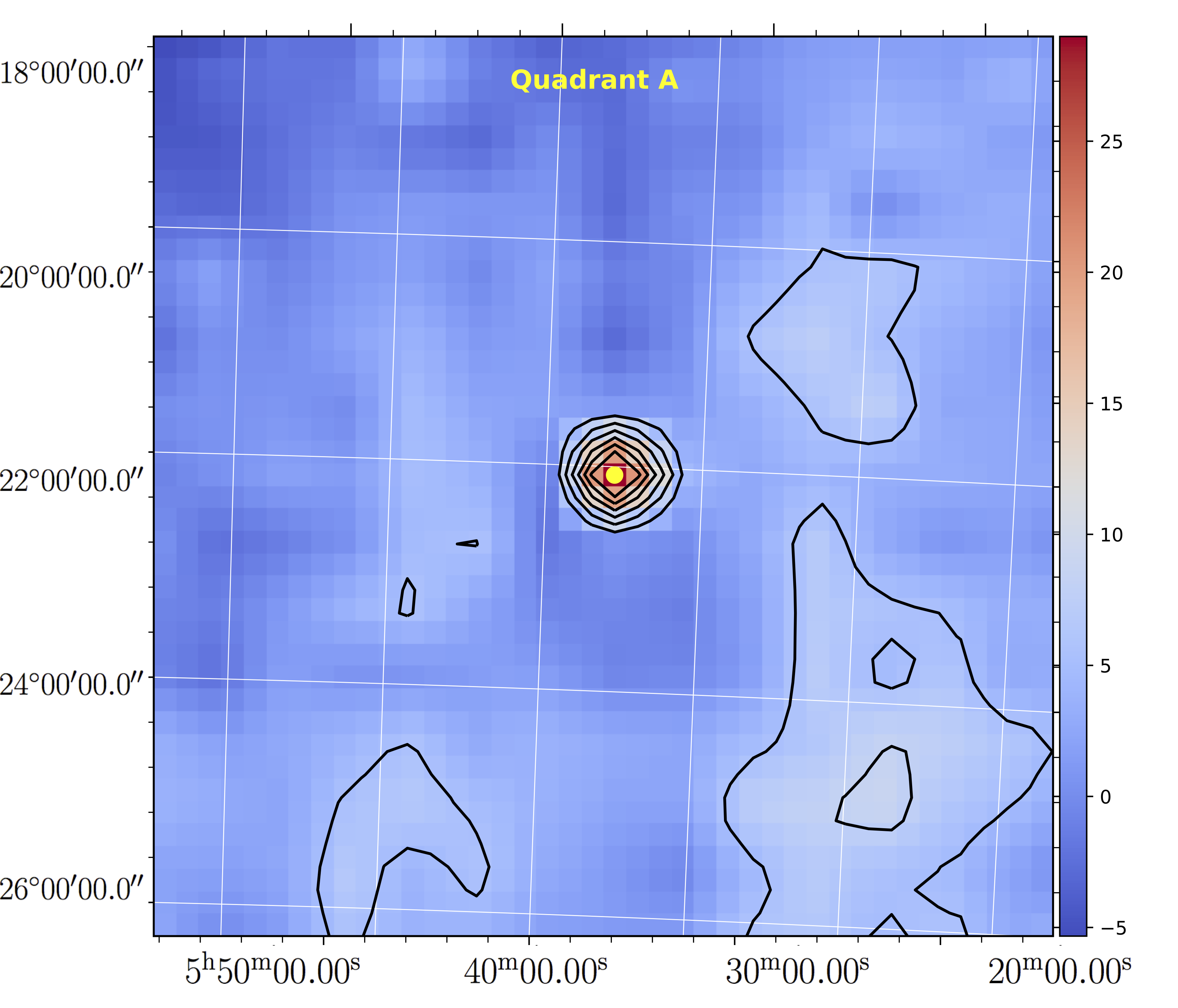}
		\caption{Quadrant A. }
	\end{subfigure}
	\begin{subfigure}{0.45\textwidth}
	    \centering
		\includegraphics[width=\textwidth]{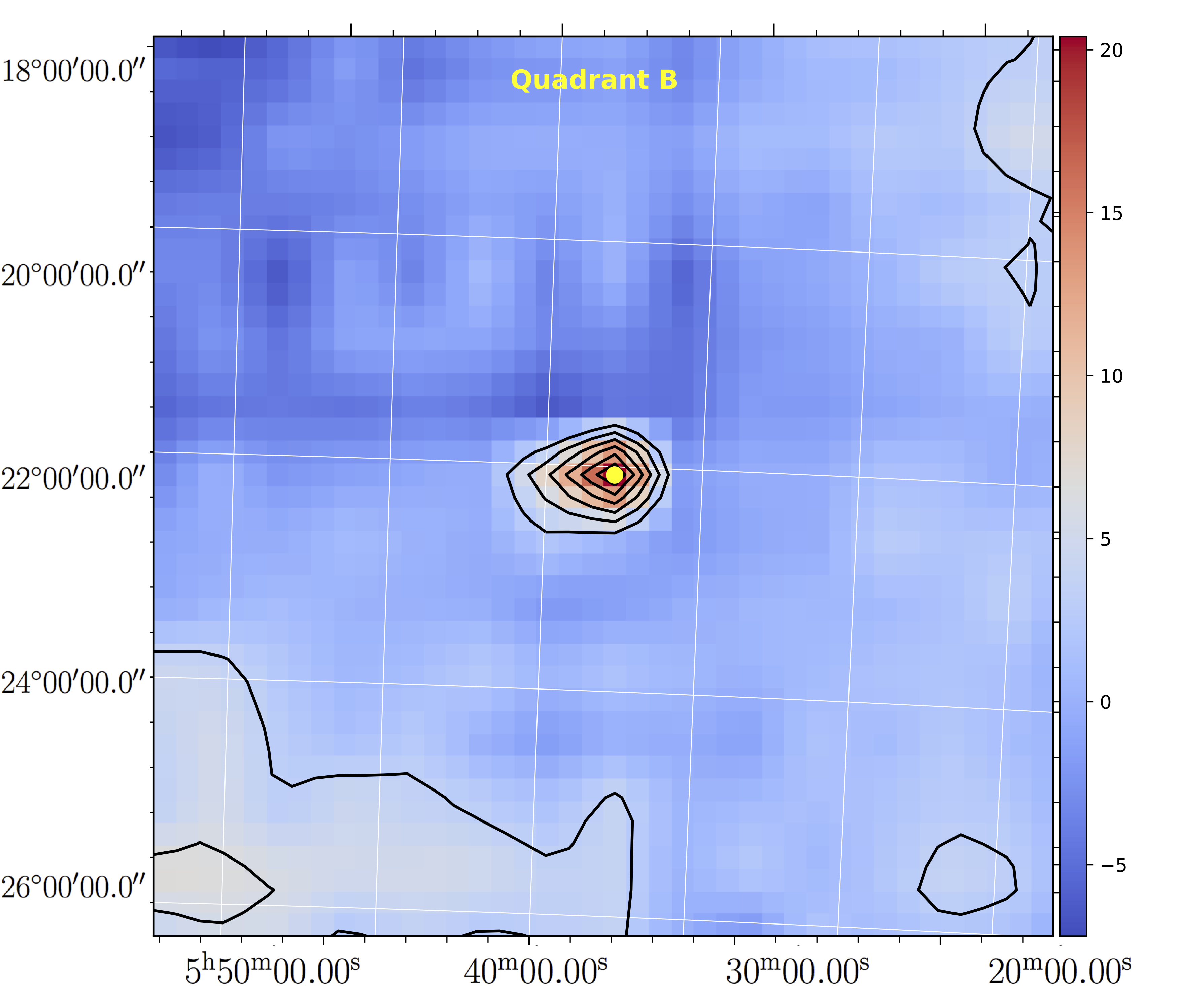}
		\caption{ Quadrant B.}
	\end{subfigure}
	\newline
	\begin{subfigure}{0.45\textwidth}
	    \centering
		\includegraphics[width=\textwidth]{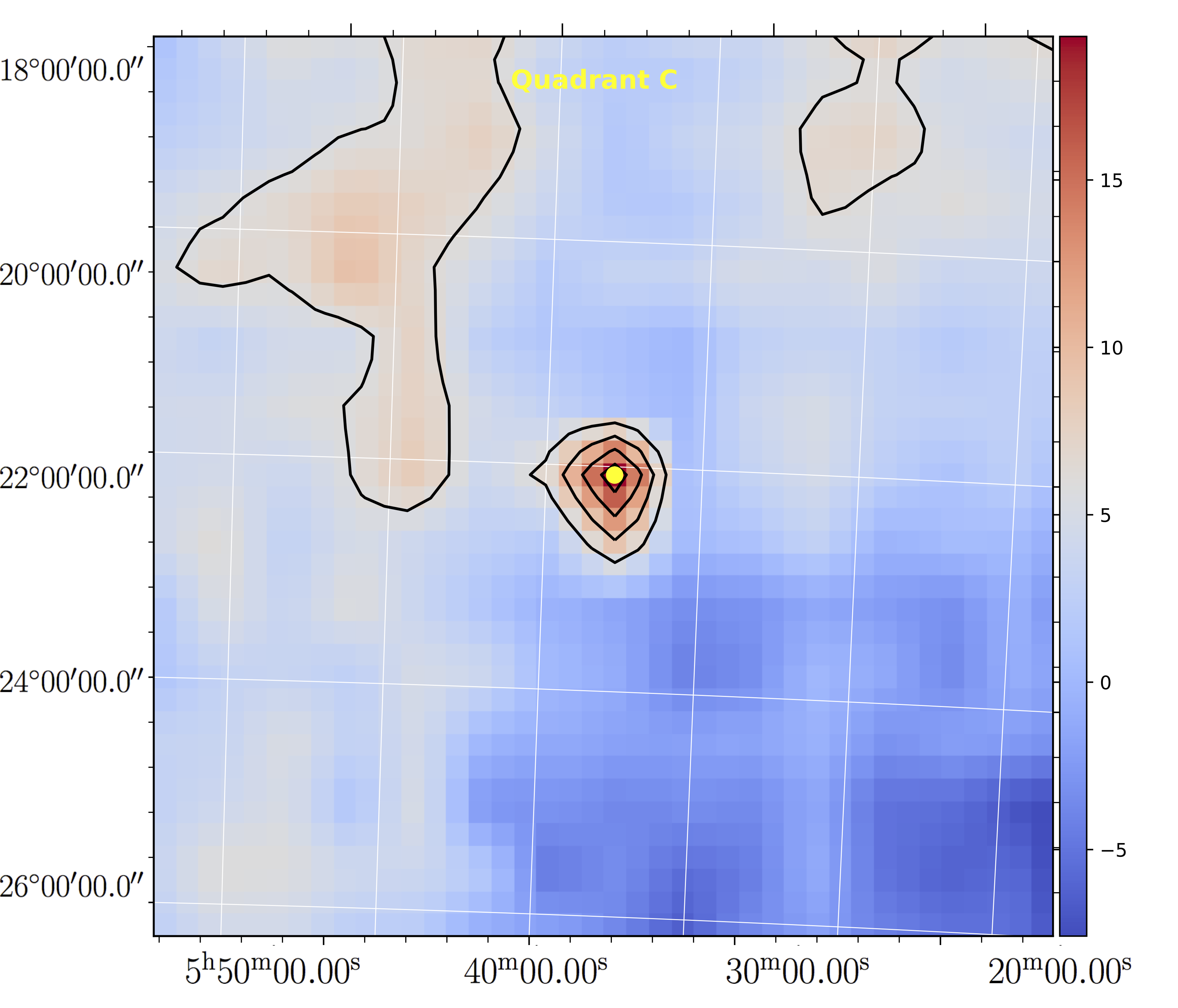} 
		\caption{Quadrant C. }
	\end{subfigure}
	\begin{subfigure}{0.45\textwidth}
	    \centering
		\includegraphics[width=\textwidth]{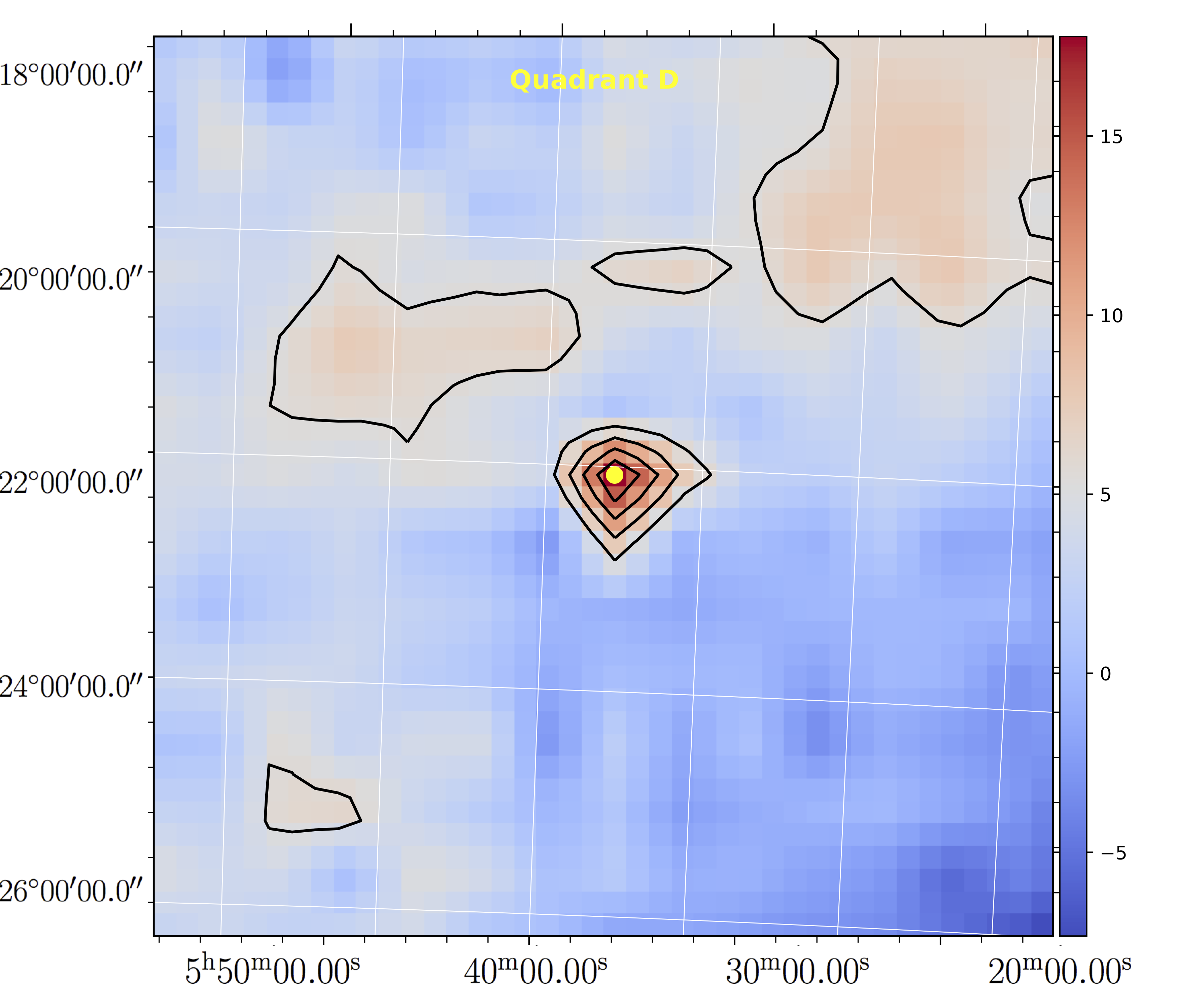} 
		\caption{Quadrant D.}
	\end{subfigure}
    	\caption[Shift corrected sky images using Balanced cross-correlation]{Reconstructed sky images using the shadow library corrected for mask shifts by Balanced cross-correlation technique}
    	\label{fig_balancedcc_image_fixed}
\end{figure*}

Figure~\ref{fig_balancedcc_image_fixed} shows the sky images reconstructed using the balanced cross-correlation. The reconstruction is performed using the shadow library corrected for the mask shift. In reconstructed images, the source is now at the expected location and also the extension in quadrant B is reduced to a great extent.

\section{Conclusions} 
We have described the on-ground calibration and in-flight calibration performed for the imaging performance of CZTI. Using the qualification model, we verified the design of the CZTI coded mask and reconstruction algorithms.  We verified that the performance of the CZTI meets the design goal of imaging resolution 8~arcmin or better. During the in-flight calibration, we found that there is a misalignment between the detector and the mask in three of its four quadrants.  The estimated mask shift with respect to the detector in quadrant B was $-1.45$~mm in the X-direction, while those in quadrants C and D were +1.68~mm and +1.50~mm respectively in the Y direction. We mitigated the effect of this misalignment by modifying the algorithms employed for image reconstruction.

\section*{Acknowledgement}

This publication uses data from the AstroSat mission of the Indian Space Research Organisation (ISRO), archived at the Indian Space Science Data Centre (ISSDC). The CZT Imager is built by a consortium of Institutes across India including Tata Institute of Fundamental Research, Mumbai, Vikram Sarabhai Space Centre, Thiruvananthapuram, ISRO Satellite Centre, Bengaluru, Inter University Centre for Astronomy and Astrophysics, Pune, Physical Research Laboratory, Ahmedabad, Space Application Centre, Ahmedabad: contributions from the vast technical team from all these institutes are gratefully acknowledged.
\bibliography{reference}

\end{document}